\documentclass[11pt]{article}
\usepackage{graphicx,subfigure}
\usepackage{textcomp}
\usepackage{amssymb}
\usepackage{amsmath}
\usepackage{fixltx2e}
\usepackage{authblk}
\usepackage{xcolor}
\newcommand{\beq}{\begin{equation}}
\newcommand{\eeq}{\end{equation}}
\newcommand{\beqa}{\begin{eqnarray}}
\newcommand{\eeqa}{\end{eqnarray}} \newcommand{\eeqan}{\end{eqnarray*}}
\newcommand{\ba}{\begin{array}}
\newcommand{\ea}{\end{array}}
\newcommand{\ben}{\begin{enumerate}}
\newcommand{\een}{\end{enumerate}}
\newcommand{\bfl}{\begin{flushleft}}
\newcommand{\efl}{\end{flushleft}}
\newcommand{\btab}{\begin{tabular}}
\newcommand{\etab}{\end{tabular}}
\newcommand{\bit}{\begin{itemize}}
\newcommand{\eit}{\end{itemize}}
\newcommand{\bdes}{\begin{description}}
\newcommand{\edes}{\end{description}}
\newcommand{\bdm}{\begin{displaymath}}
\newcommand{\edm}{\end{displaymath}}

\newcommand{\no}{\nonumber}

\newcommand{\dg}{\dagger}

\newcommand{\cL}{{\cal L}}

\newcommand{\del}{\partial}

\title{The reactions $\pi\pi\rightarrow\pi\pi$ and $\gamma\gamma\rightarrow\pi\pi$
in $\chi$PT with an isosinglet scalar resonance}
\author[1,2]{Arbin Thapaliya \footnote{Email: AThapaliya@franklincollege.edu}}
\author[2]{Daniel R. Phillips \footnote{Email: phillid1@ohio.edu}}
\affil[1]{Department of Chemistry and Physics, Franklin College, Franklin, Indiana 46131 USA}
\affil[2]{Institute of Nuclear and Particle Physics and Department of
Physics and Astronomy,

Ohio University, Athens, Ohio 45701 USA}
\date{}                     
\begin{document}
\maketitle

\begin{abstract}
The lowest-lying resonance in the QCD spectrum is the $0^{++}$ isoscalar $\sigma$ meson, also known as the $f_0(500)$. 
We augment SU(2) chiral perturbation theory ($\chi$PT) by including
the $\sigma$ meson as an additional explicit degree of freedom, as proposed by Soto, Talavera, and Tarr\'us and others.
In this effective field theory, denoted $\chi$PT$_S$, the $\sigma$ meson's well-established mass and decay width are not sufficient to 
properly renormalize its self energy. At $\mathcal{O}(p^4)$ another low-energy constant appears in the 
dressed $\sigma$-meson propagator; we adjust it so that the isoscalar pion-pion scattering length is also reproduced. 
We compare the resulting amplitudes for the $\pi\pi\rightarrow\pi\pi$ and $\gamma\gamma\rightarrow\pi\pi$
reactions to data  from threshold through 
the energies at which the $\sigma$-meson resonance affects
observables.
The leading-order (LO) $\pi \pi$ amplitude reproduces the $\sigma$-meson pole position, the isoscalar $\pi \pi$ scattering lengths and 
$\pi \pi$ scattering and $\gamma \gamma \rightarrow \pi \pi$ data  up to 
$\sqrt{s} \approx 0.5$ GeV. It also yields a $\gamma\gamma\rightarrow\pi\pi$ amplitude that obeys the Ward identity. 
The value obtained for the $\pi^0$ polarizability is, however, only slightly larger than that obtained in standard $\chi$PT.
\end{abstract}
\section{Introduction}

The spectrum of Quantum Chromodynamics (QCD) consists of several bound and resonant states with masses below 1 GeV. The lightest
QCD bound states are the pseudoscalar pions, which have a special role in the theory as pseudo-Nambu-Goldstone bosons of QCD's approximate, 
spontaneously-broken, chiral symmetry.  
The lowest-lying QCD resonance has $0^{++}$ quantum numbers: the same as those of the vacuum. This state, often termed the ``$\sigma$ meson", and also referred to as the $f_0(500)$, is (slightly) manifested in pion-pion scattering~ \cite{Pelaez:2015qba}. It has attracted much attention
over many years---indeed the suggestion that the meson spectrum contains a somewhat light scalar pre-dates QCD itself~\cite{johnson,schwinger}. We do not review that history further here, but instead refer to Ref.~\cite{Pelaez:2015qba} for a summary and further references. 

Determinations of $\sigma$-meson parameters rely on an extrapolation of the $\pi \pi$ scattering amplitude into the complex plane: one must 
obtain its mass, $M_\sigma$, and 
width, $\Gamma$, from the position in the complex energy plane at which a pole in the $\pi \pi$ $t$-matrix occurs. 
From 1996--2010 the Particle Data Group (PDG) \cite{Agashe:2014kda} results for the mass and decay width ranged
from 400 to 1200 MeV and from 500 to 1000 MeV, respectively.
These wide variations occurred because obtaining the mass, decay width, and couplings of this resonance
is difficult: the resonance is very broad and can hardly be seen in the $\pi\pi$ scattering
phase shifts. The standard Breit-Wigner formulation for narrow resonances is definitely not applicable
in this case. The last fifteen years has seen the advent of dispersion-relation evaluations that incorporate the constraints
of chiral symmetry and---in some cases---crossing symmetry too~\cite{Caprini:2005zr,Colangelo:2001df,GarciaMartin:2011jx,
Moussallam:2011zg}. The results of these calculations largely agree, and the 2015 review of Pel\'aez quotes a $\sigma$-meson pole position~\cite{Pelaez:2015qba}:
\begin{equation}
\sqrt{s}=M_\sigma - i \Gamma/2; \quad M_\sigma=(449^{+22}_{-16})~{\rm MeV}; \quad \Gamma=(550 \pm 24)~{\rm MeV}.
\label{eq:sigmapole}
\end{equation}

This result implies that QCD's spectrum includes a scalar isosinglet state at low mass. That is in accord with a recent lattice QCD calculation by Brice\~{n}o and collaborators~\cite{Briceno:2016mjc}. They find that at pion masses 
$m_\pi \approx 400$ MeV the $\sigma$ is a $\pi \pi$ bound state, but, as the quark mass in their simulation is lowered (ultimately to a smallest value of $m_\pi=236$ MeV), this state
becomes a broad resonance. 

The pole position (\ref{eq:sigmapole}) is markedly lower than the scale of chiral-symmetry breaking, $\Lambda_{\chi {\rm SB}}$, which is usually understood to be the rho-meson mass, or $4 \pi F$, with $F=92.419$ MeV the pion decay constant. It is also comparable to the kaon mass. This has led some authors to propose that the $\sigma$ is itself a (pseudo-)Goldstone boson of QCD. Crewther and Tunstall developed an effective field theory (EFT) based on the postulated existence of a non-perturbative infra-red fixed point in the flow of the strong coupling constant $\alpha_S$ in three-flavor QCD, and the consequent emergence of a QCD dilation which they identified with the $\sigma$~\cite{Crewther:2015dpa}. The resulting ``chiral-scale perturbation theory" includes as dynamical degrees of freedom the eight Goldstone bosons of SU(3) chiral perturbation theory and the $\sigma$. In fact, for a sufficiently large number of flavors, $N_f$, QCD can be expected to develop an infra-red fixed point and hence a conformal symmetry at long distances. Recent lattice studies with $N_f=8$ support this expectation~\cite{Aoki:2014oha,Appelquist:2016viq}. However, in such a theory there is neither confinement nor chiral-symmetry breaking. Golterman and Shamir  examined an extension of QCD with $N_f$ large enough that the theory is on the verge of developing an infra-red fixed point, but not so large that the theory ceases to display confinement and chiral symmetry breaking. They conjectured that dilatation symmetry is recovered in a triple limit: the chiral limit of massless quarks, the large-$N_c$ limit (with $N_f/N_c$ held fixed), and the limit that the number of flavors approaches the critical value for conformality. They then developed a low-energy EFT for the pions and the dilatonic meson by making an expansion in the three small parameters associated with these different aspects of conformal symmetry breaking~\cite{Golterman:2016lsd} (cf. the more recent Ref.~\cite{Appelquist:2017wcg}). 

However, it is not necessary to assume that the $\sigma$ is an (approximate) QCD dilaton in order to include it as an explicit degree of freedom in the low-energy EFT. After all, that $|M_\sigma - i \Gamma/2|$ is well below $\Lambda_{\chi {\rm SB}}$ is an empirical fact. This $0^{++}$ resonance can therefore be expected to spoil the convergence of any perturbative expansion in channels where it plays a role (cf. Ref.~\cite{Cecile:2008kp} for lattice studies of QCD-like theories where this clearly happens).
This motivates augmenting standard chiral perturbation theory by the addition of a $\sigma$ field, whose mass is midway between the pseduo-Goldstone-boson mass scale, $m_\pi$, and $\Lambda_{\chi {\rm SB}}$. The resulting EFT has a (spontaneously and dynamically broken) SU(2)$_L$ $\times$ SU(2)$_R$ symmetry. It was written down by Soto, Talavera, and Tarr\'us in Ref.~\cite{Soto:2011ap}, who called it $\chi$PT$_S$. $\chi$PT$_S$ has also been explored by Ametller and Talavera~\cite{Ametller:2014vba,Ametller:2015xva} and Hansen {\it et al.}~\cite{Hansen:2016fri}. The price to be paid for not having the $\sigma$ be a Goldstone boson of a QCD symmetry is that its couplings must be fixed from data: only a few are constrained by the  chiral symmetry of the EFT. In contrast, in the approaches discussed in the previous paragraph many of the $\sigma$'s couplings are fixed. However, whether those symmetry relations prevail in nature is unclear as the connection between the version of QCD we observe experimentally and the ones considered by Golterman and Shamir and Crewther and Tunstall could be regarded as tenuous. 

Of course, the $\sigma$ was already included---together with the pions---as an explicit degree of freedom in the
 linear $\sigma$ model of Gell-Mann and Levy~\cite{GellMann:1960np}. This model reproduces many of the features of QCD's low-energy dynamics in the $0^{++}$ channel, see, e.g., Refs.~\cite{Achasov:1994iu,Achasov:2007fz}, but it is a model, rather than a systematic EFT. 
$\chi$PT$_S$ is a low-energy EFT that includes as explicit degrees of freedom the pions and a scalar. It includes a systematic expansion in a small parameter, with the Lagranian incorporating all possible operators up to a given order in that expansion. The generality of the Lagrangian means that the linear $\sigma$ model can be recovered as a special case of $\chi$PT$_S$, as explained by Soto {\it et al.}~\cite{Soto:2011ap} and Hansen {\it et al.}~\cite{Hansen:2016fri}. Specifically, $\chi$PT$_S$ does not assume that there is a LO relation between $M_\sigma$ and $F$, or (equivalently) a pre-determined value of the $\sigma \pi \pi$ coupling; since it is an EFT, $\chi$PT$_S$ makes no assumptions about the nature of the physics that generates chiral symmetry breaking at the scale $\Lambda_{\chi {\rm SB}}$. 

In this paper we examine the reactions $\pi\pi\rightarrow\pi\pi$ and $\gamma\gamma\rightarrow\pi\pi$, both of which couple to the $0^{++}$ channel in the $s$-channel, and both of which 
 exhibit slow convergence when investigated in standard, two-flavor, $\chi$PT. We compare those standard $\chi$PT calculations at leading [$\mathcal{O}(p^2)$] order to  $\chi$PT$_S$ at LO: the theory with the additional scalar isoscalar degree of freedom intercalates between $\chi$PT at $\mathcal{O}(p^2)$ and $\chi$PT at $\mathcal{O}(p^4)$. We demonstrate that $\chi$PT$_S$ naturally includes a $\sigma$ meson with a large width that is, nonetheless, not prominent in the $\pi \pi$ S-wave phase shift.
 
Our emphasis on scattering processes takes us beyond the static properties considered in Ref.~\cite{Soto:2011ap,Hansen:2016fri}. The main goal in Ref.~\cite{Soto:2011ap} was to improve extrapolations of lattice data as a function of $m_\pi$ for quantities that couple to vacuum quantum numbers. (See also the more recent Ref.~\cite{Bruns:2016}.) Both Refs.~\cite{Soto:2011ap,Hansen:2016fri} computed the corrections to the pion mass and decay constant, as well as the one-loop piece of the $\sigma$ mass (and hence the leading contribution to the $\sigma$ width) in $\chi$PT$_S$.
 
 The $\gamma \gamma \rightarrow \pi \pi$ reaction and the pion vector form factor were considered by Ametller and Talavera in Ref.~\cite{Ametller:2014vba,Ametller:2015xva}. But, as we discuss further below, they (implicitly) had a different $\sigma \pi \pi$ coupling governing the width of the $\sigma$ and its decay to two pions. This allowed Ametller and Talavera to accommodate the weak impact of the $\sigma$ on the $\gamma \gamma \rightarrow \pi \pi$ cross section, yet also incorporate a $\sigma$ with a large width in their theory. But such a treatment is both inconsistent and unnecessary: we will show below that a proper treatment of the reaction in $\chi$PT$_S$ obeys the Ward identity and does not require this inconsistency in the $\pi \pi$ amplitude.
    
Our approach also differs from these previous works in that we employ a power counting with two light scales: $m_\pi$ and $M_\sigma$. The resulting hierarchy on which the EFT is built is then $m_\pi \ll M_\sigma \ll \Lambda_{\chi {\rm SB}}$. A particular virtue of this hierarchy is that the loop effects that generate the $\sigma$ width in the $s$-channel are perturbative for values of Mandelstam $s$ that are $\sim m_\pi^2$, i.e., within the purview of the EFT but not close to $M_\sigma^2$. However, for $s \sim M_\sigma^2$ the infra-red singularity in the (nominal) LO $\sigma$ propagator mandates the resummation of the one-loop self energy, thereby generating a width for the resonance. For the processes that we consider it is always the case that the $\sigma$ pole in the $t$- and $u$-channels is far away, so $t$- and $u$-channel $\sigma$ exchanges are higher order in the $\chi$PT$_S$ expansion. Thus the LO amplitude in our approach consists of the standard $\chi$PT $\mathcal{O}(p^2)$ interaction plus an $s$-channel $\sigma$ pole that is enhanced near the resonance so it becomes $\mathcal{O}(p^0)$. (Away from $s \sim M_\sigma^2$ the $s$-channel pole is $\mathcal{O}(p^4/M_\sigma^2)$.) 
This LO amplitude does violate crossing symmetry, but
it does so only by corrections that are perturbative both for $s \sim M_\sigma^2$ and in the near- and sub-threshold region where $s \sim m_\pi^2$. This same three-scale strategy has been successfully employed for the $\Delta(1232)$ resonance in the low-energy EFT of the single-baryon sector~\cite{Pascalutsa:2003,Pascalutsa:2006up,McGovern:2012ew}. 

The rest of this paper is structured as follows: in Sec.~\ref{sec:Lagpc} we review the Lagrangian developed in Ref.~\cite{Soto:2011ap} (or, equivalently, the later Ref.~\cite{Hansen:2016fri}), and explain the power counting we use in this paper. In Sec.~\ref{sec:sigmaSigma} we calculate the $\sigma$ propagator at $\mathcal{O}(p^4)$. In Sec.~\ref{sec:pipi} we employ this propagator, together with the standard mechanisms of $\chi$PT at $\mathcal{O}(p^2)$, to describe $\pi \pi$ scattering. In Sec.~\ref{sec:gammagammapipi} we consider $\gamma \gamma \rightarrow \pi \pi$. We first discuss the Ward identity for this reaction, and also explain why it is important to have a consistent treatment of the $\sigma$ width and the $\sigma \rightarrow \pi \pi$ vertex---something that was not achieved in Ref.~\cite{Ametller:2014vba,Ametller:2015xva}. We then show that the good description of phase shifts in Sec.~\ref{sec:pipi} carries over to a nice reproduction of the $\gamma \gamma \rightarrow \pi^0 \pi^0$ cross section up to $\sqrt{s} \approx 600$ MeV.  However, the relatively weak impact of the $\sigma$ meson in this process also means it produces only a small increase in the pion's dipole polarizability. This is in contrast to, e.g., the Roy equation approach of Ref.~\cite{Hoferichter:2011wk} where the $\sigma$ pole accounts for about half of the difference between the $\mathcal{O}(p^4)$ and $\mathcal{O}(p^6)$  numbers for the $\pi^0$'s dipole polarizability. Sec.~\ref{sec:conclusion} offers our conclusions.
 
\section{The Lagrangian and Power counting}

\label{sec:Lagpc}

In Ref.~\cite{Soto:2011ap} Soto {\it et al.} modified the $\chi$PT Lagrangian---which is approximately invariant under SU(2)$_L$ $\times$ SU(2)$_R$ 
transformations---to $\chi$PT$_S$ by including the terms containing an isosinglet scalar $\sigma$ field $S$. (See also Ref.~\cite{Ecker:1988te} for a chirally-symmetric Lagrangian that incorporates an explicit scalar degree of freedom.) $S$ is then a dynamical degree of freedom that 
is in addition to the matrix $U$ that parameterizes the Goldstone boson fields in standard SU(2) $\chi$PT~\cite{Gasser:1982ap,Gasser:1983yg,Scherer:2012xha}. In the notation of Ref.~\cite{Soto:2011ap} the terms
in the effective Lagrangian that are relevant for this study are:
\beqa
\cL_2^S & = & \bigg(\frac{F^2}{4} + F c_{1d} S  + c_{2d} S^2 + \cdots \bigg)   \langle D_\mu U (D^\mu U)^\dg\rangle
+ \left(\frac{F^2}{4} + c_{1m} S + c_{2m} S^2 + \ldots \right) \langle\chi U^\dagger+U\chi^\dagger\rangle
\no\\*&&~~~~~~~~~~~~+ \frac{1}{2}\del_\mu S\del^\mu S-\frac{1}{2}m_S^2SS - f_{2p} (\del_\mu\del^\mu S)^2 - \frac{\lambda_3}{3!} S^3 - \frac{\lambda_4}{4!} S^4~,
\label{eq1}
\eeqa
where $c_{1d}$, $c_{2d}$, $c_{1m}$, $c_{2m}$, and $f_{2p}$ are new low-energy constants (LECs) in the $\mathcal{O}(p^2)$ Lagrangian. 
In Eq.~(\ref{eq1}), $D_\mu U=\partial_\mu U - i [v_\mu,U] + i \{a_\mu,U\}$ is the chiral covariant derivative, $\chi$ represents a scalar source (and hence is where 
quark masses enter the theory)
 and 
the symbol $\langle \cdots \rangle$ is
the isospin trace of the matrix within it. 
The terms on the second line are the Lagrangian of the scalar $S$ field with bare mass $m_S$. This Lagrangian, unlike
the typical ${\cal L}$ for scalar fields, contains an additional fourth-order term which is Lorentz invariant.  Although this term appears only in the $\mathcal{O}(p^4$) Lagrangian it is needed for proper renormalization
of the $\sigma$-meson self energy. In the absence of this term,
the $\sigma$ couples strongly to two pions, causing
the bumps in the $\pi\pi\rightarrow\pi\pi$ and $\gamma\gamma\rightarrow\pi\pi$ processes to be very pronounced---something that is not seen in data.

The LEC $c_{1m}$ must be zero at tree level (and must be additionally tuned at loop level) to stop the scalar field $S$ mixing with the vacuum~\cite{Soto:2011ap}. Soto {\it et al.} also take $\lambda_3=\lambda_4=0$, in order to implement the (presumed) triviality of strongly-coupled scalar field theories in four dimensions in the EFT. Once the scalar is coupled to Goldstone bosons (e.g., through the couplings $c_{1d}$, $c_{2d}$, and $c_{2m}$) values of $\lambda_3$ and $\lambda_4$ of $\mathcal{O}(m_S^2/\Lambda_{\chi {\rm SB}})$ and $\mathcal{O}(m_S^2/\Lambda_{\chi {\rm SB}}^2)$ will be induced through renormalization of pion-loop contributions to the three- and four-scalar correlation functions. This, however, only affects Goldstone-boson reactions beyond the order to which we work here.

In $\chi$PT$_S$ we consider two different energy regions of interest. In the near-threshold region we have $p \sim m_\pi$ and the standard $\chi$PT power counting:
each vertex with $n$ powers of momentum $p$ or $m_\pi$ scales as $p^{n}$ 
and the pion propagator scales as
$p^{-2}$. In this regime the $\sigma$ propagator scales as $m_S^{-2}$, since $p$ is markedly less than $m_S$. It therefore produces larger threshold effects than the $\chi$PT counter terms at $\mathcal{O}(p^4)$, since $m_S$ is taken to be $\ll \Lambda_{\chi {\rm SB}}$. 

But the effects of the $\sigma$ are enhanced---to an effect that is nominally larger than the $\mathcal{O}(p^2)$ leading $\chi$PT $\pi \pi$ amplitude---in the second regime where $p \sim m_S$, i.e., in the vicinity of the resonance. Here the $\sigma$ propagator develops a pole. It then needs to be dressed by the inclusion of the leading $[\mathcal{O}(p^4)]$ self energy, $\Sigma$, which is resumed to all orders in the $s$-channel via a Dyson equation. The inclusion of $f_{2p}$ as part of this self energy is mandatory for proper renormalization, which is why this particular $p^4$ operator is relevant in our leading-order study. The renormalized $\Sigma$ ensures that the $\sigma$ develops a pole at the physical mass and width. There is then a (in principle narrow) kinematic window where the $p^2 - M_\sigma^2$ piece of the inverse $\sigma$ propagator is of the same order, or smaller than, the $\mathcal{O}(p^4)$ self energy. In this kinematic window the resumed $\sigma$ propagator scales as $p^{-4}$, is enhanced, and becomes a leading-order effect.

We note that in this second, near-resonance, kinematic domain, vertices proportional to $m_\pi^2$ are suppressed compared to vertices proportional to $p^2$. For example, the effect of $c_{2m}$ in this region is suppressed by factors of $m_\pi^2/M_\sigma^2$ compared to corrections to the propagator proportional to $s$. For this reason, in what follows, we do not consider pieces of the $\sigma$ self energy that involve two vertices each $\sim m_\pi^2$. 
The details of the renormalization of the one-loop $\sigma$ self energy by terms involving two insertions of the chiral-symmetry-breaking quantity $\chi$ that were worked out in Ref.~\cite{Soto:2011ap} are a higher-order effect in our approach. 

\section{Calculation of the $\sigma$-meson self energy}

\label{sec:sigmaSigma}

\begin{figure}[h]
\centering
 \includegraphics[scale=0.3]{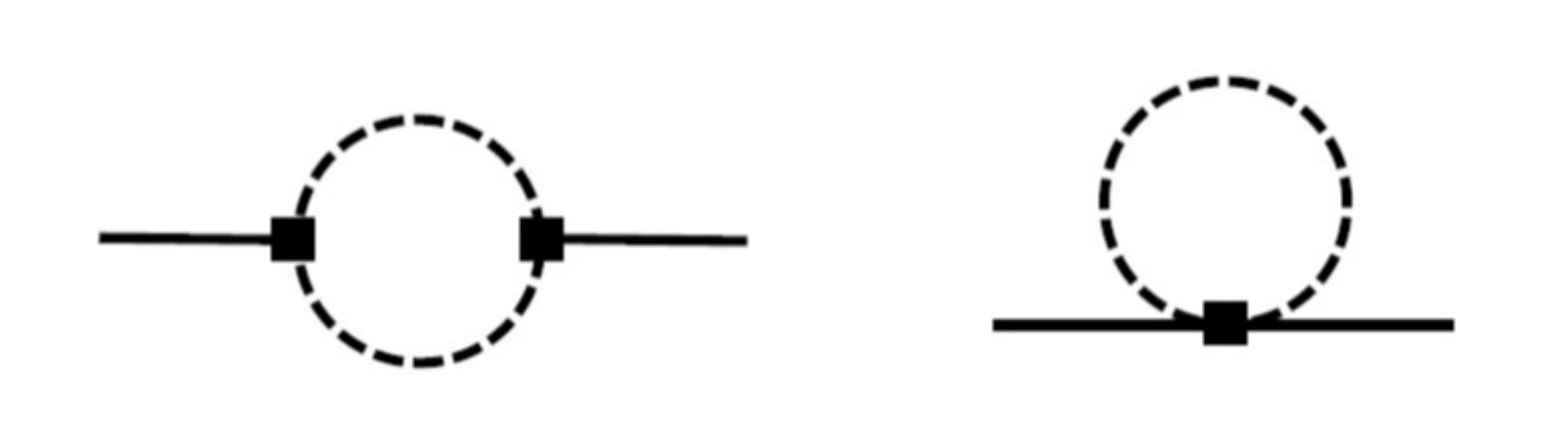}
\caption{The self-energy diagrams of the $\sigma$ meson at the one-pion-loop level, i.e. $\mathcal{O}(p^4)$. The solid (dashed) line represents
the $\sigma$($\pi$) propagator. The squares indicate that the interaction appears
due to the terms containing the $S$ field in the $\cL_2^S$ Lagrangian. The left-hand diagram involves two insertions of $c_{1d}$ and the right-hand one one insertion of $c_{2d}$ or $c_{2m}$.}
\label{ppfig2a}
\end{figure}
We now perform calculation of the $\sigma$-meson self energy to 
$\mathcal{O}(p^4$). Only one-loop diagrams need to be considered and the pertinent ones are 
shown in Fig.~\ref{ppfig2a}.  
We can express the 
$\sigma$ self energy in the modified minimal-subtraction ($\overline{MS}$) renormalization scheme as 
\beqa
\label{alpha1}
\Sigma^{\overline{MS}}(s,\overline{\mu})&=&\Sigma_0(\overline{\mu})+\Sigma_1(\overline{\mu})s+\Sigma_2(\overline{\mu})s^2
+c_{1d}^2(\overline{\mu})\tilde{\Sigma}(s)~,
\eeqa
with
\beqa
\label{alpha2}
\Sigma_0(\overline{\mu})&=&-\frac{3m_\pi^4}{2\pi^2F^2}\bigg[c_{1d}^2\bigg\{2+3\ln\overline{\mu}^2\bigg\}+(c_{2d}-c_{2m})\bigg\{1+\ln\overline{\mu}^2-\ln m_\pi^2\bigg\} \bigg]~,\no\\*
\Sigma_1(\overline{\mu})&=&-\frac{3c_{1d}^2}{2\pi^2F^2}\int_0^1dx\bigg[-2x(1-x)m_\pi^2\bigg\{2+3\ln\overline{\mu}^2\bigg\}\no\\*
&&~~~+\bigg\{\frac{1}{4}(1-2x)^2-2x(1-x)\bigg\}(-m_\pi^2)\bigg\{-1-2\ln\overline{\mu}^2)\bigg\}\Bigg]~,\no\\*
\Sigma_2(\overline{\mu})&=&-\frac{3c_{1d}^2}{2\pi^2F^2}\int_0^1dx\bigg[x^2(1-x)^2\bigg\{2+3\ln\overline{\mu}^2\bigg\}\no\\*
&&~~~+\bigg\{\frac{1}{4}(1-2x)^2-2x(1-x)\bigg\}x(1-x)\bigg\{-1-2\ln\overline{\mu}^2\bigg\}+x^2(1-x)^2\ln\overline{\mu}^2\Bigg]~,\no\\*
\tilde{\Sigma}(s)&=&-\frac{3}{2\pi^2F^2}\int_0^1dx\bigg[3D^2-2\bigg\{\frac{1}{4}(1-2x)^2-2x(1-x)\bigg\}sD+x^2(1-x)^2s^2\Bigg](-\ln(-D))~,\no\\* 
\eeqa
where $\overline{\mu}$
is a renormalization scale and $D=x(1-x)s-m_\pi^2+i\epsilon$.

In Eq.~(\ref{alpha2}), $\Sigma_0$, $\Sigma_1$, and $\Sigma_2$ are $\overline{\mu}$-dependent 
whereas $\tilde{\Sigma}(s)$ is independent of $\overline{\mu}$. When
Eq.~(\ref{alpha1}) is combined with bare propagators and vertices
the $\Sigma_0(\overline{\mu})$-term is renormalized by $m_S^2(\overline{\mu})$ and the term linear in $s$
by $c_{1d}^2(\overline{\mu})$. However, the left-hand graph in Fig.~\ref{ppfig2a} is quartically divergent, and so there is also
an $s^2 \ln(\overline{\mu})$ piece of the diagram that must be absorbed by a counterterm. This is done by $f_{2p}(\overline{\mu})$. 
We then express the 
dressed renormalized $\sigma$ propagator as
\beqa
iD(s)&=&\frac{i}{s-m_{S,r}^2-2 f_{2p,r} s^2-c_{1d,r}^2\tilde{\Sigma}(s)}~,   
\label{alpha4}
\eeqa
where the quantities with subscripts $r$
are the $\overline{\mu}$-independent renormalized quantities. 
We note that Bruns has also recently computed the $\sigma$-meson self energy, and observed the presence of the $s^2 \ln(\mu)$ term that we found here~\cite{Bruns:2016}. However, he then expands the propagator
around the pole, and argues that the quadratic-in-$s$ part is irrelevant for his results. In what follows we keep $f_{2p,r}$ as a free parameter in our calculation.

Equation (\ref{alpha4}) shows that there are three unknown parameters---$m_{S,r}$, $f_{2p,r}$, and $c_{1d,r}$---that affect the $\sigma$-meson physics in $\chi$PT$_S$. 
Two constraints on them are obtained by demanding that the quadratic $s$-dependence $\sim f_{2p,r}$ and the $s$-dependence of $\tilde{\Sigma}(s)$ in
Eq.~(\ref{alpha4})
ultimately produce a pole at the position (\ref{eq:sigmapole}):
\beqa
M_\sigma^2 -\frac{\Gamma^2}{4}-m_{S,r}^2-2f_{2p,r}\Bigg(M_\sigma^4-6M_\sigma^2\frac{\Gamma^2}{4}
+\frac{\Gamma^4}{16}\Bigg)+c_{1d,r}^2\Re[\tilde{\Sigma}([M_\sigma - i \Gamma/2]^2)]& = & 0~,
\no\\*M_\sigma\Gamma+2f_{2p,r}\Bigg(4M_\sigma\frac{\Gamma^3}{8}-4M_\sigma^3\frac{\Gamma}{2}\Bigg)
-c_{1d,r}^2\Im[\tilde{\Sigma}([M_\sigma - i\Gamma/2]^2)]& = & 0~,
\label{NEWeq1}
\eeqa
where $\Re$ and $\Im$ denote the real and imaginary parts of $\tilde{\Sigma}$. Note that the pole is not at $s=m_{S,r}^2$. 
The third constraint results from demanding that the LO amplitude reproduce the experimental pion-pion scattering length in the scalar-isoscalar channel,
$a_0^0$=$0.2210(47)(40)~m_\pi^{-1}$~\cite{Batley:2010}. (For details on obtaining the $\pi \pi$ amplitude that yields this scattering length from the propagator (\ref{alpha4})
see Sec.~\ref{sec:pipi} below.) 
Here, and throughout, we take $F=92.419$ MeV and $m_\pi=139.57$ MeV~\cite{Agashe:2014kda}. 
The values of $m_{S,r}$, $f_{2p,r}$, and $c_{1d,r}$ that we then obtain are
\begin{equation}
m_{S,r}= 221^{+5}_{-4}~{\rm MeV}; \quad c_{1d,r}=0.206^{+0.001}_{-0.002}; \quad
f_{2p,r}=(3.4^{+0.01}_{-0.02}) \times 10^{-6}~{\rm MeV}^{-2}.
\label{eq:parameters}
\end{equation}
These are the $\chi$PT$_S$ parameters for the particular set of renormalization conditions we chose for the leading-order amplitude: other choices of
renormalization condition are certainly possible.

The errors in Eq.~(\ref{eq:parameters}) result solely from propagation of the uncertainties in the data, and do not account for the impact that higher-order corrections might have on
these parameters. While the pole position will not
change, the determination of $c_{1d,r}$ could be affected by the appearance of other $\sigma \pi \pi$ couplings at higher orders in the EFT expansion, e.g., 
those proportional to the quark mass. This, in turn, will alter the balance between the different terms in the denominator of Eq.~(\ref{alpha4}) and hence the values
of $f_{2p,r}$ and $m_{S,r}$. Moreover, the values of $c_{1d,r}$, $m_{S,r}$, and $f_{2p,r}$ obtained at higher order 
will also change as new graphs enter the $\pi \pi$ scattering amplitude. Some of these
higher-order contributions are discussed in Sec.~\ref{sec:pipi} below. However, the parametric suppression of higher-order corrections
in $\chi$PT$_S$ implies that the determination (\ref{eq:parameters}) should be accurate up to a relative error $\sim M_\sigma^2/\Lambda_{\chi {\rm SB}}^2$.
This should also be the largest possible size of the shift in the numbers if different renormalization conditions are employed. 

Two points must be noted in comparing our results to those of Soto {\it et al.} in Ref.~\cite{Soto:2011ap}. First, Soto {\it et al.} pointed out the need for renormalization of the one-loop $\sigma$ self energy, but they set the finite part of $f_{2p}$ to
zero. (This is ultimately equivalent to Bruns removing the $s^2$ piece from the propagator he considers~\cite{Bruns:2016}.)
We find a non-zero, but natural, value:  $f_{2p,r} \sim 1/\Lambda_{\chi {\rm SB}}^2$.

Second, our analytic result for the self energy $\tilde{\Sigma}$ agrees with that found in Ref.~\cite{Soto:2011ap}. However, Soto {\it et al.} took 
\begin{equation}
\frac{\Gamma}{2}=\frac{c_{1d,r}^2}{M_\sigma} \Im\tilde{\Sigma}(M_\sigma^2).
\label{eq:factoroftwo}
\end{equation}
The key difference to our Eq.~(\ref{NEWeq1}) is that this relation between the width and self energy includes an incorrect factor of two in the denominator on the left-hand side. 
It is true that Soto {\it et al.} also evaluated the self energy for real $s$ to obtain their width, i.e., they treated $\Gamma$ as a perturbative correction to the LO mass. 
They also, as already noted, took $f_{2p,r}=0$. However, both of these are consistent with our result (\ref{NEWeq1}) in appropriate limits. This factor of two in Eq.~(\ref{eq:factoroftwo}) is not. 
Our result for $\Gamma$ in the same limit that Soto {\it et al.} considered is:
\begin{equation}
\Gamma=\frac{3 c_{1d,r}^2}{8 \pi F^2 M_\sigma} \sqrt{1 - \frac{4 m_\pi^2}{M_\sigma^2}}(M_\sigma^2 - 2 m_\pi^2)^2,
\label{eq:ouranalyticGamma}
\end{equation}
although we stress that this is not the width we evaluate since we solve Eq.~(\ref{NEWeq1}) for complex values of $s$ on the second Riemann sheet. 
 Hansen {\it et al.} state
they reproduce the result of Soto {\it et al.} for the $\sigma$ width in a particular limit of their (more general) calculation. 
Consequently the analytic formula (\ref{eq:ouranalyticGamma}) is also a factor of two smaller 
than that of Ref.~\cite{Hansen:2016fri}. 

Indeed, if we adopt the same strategy as Ref.~\cite{Soto:2011ap,Bruns:2016} and set $f_{2p,r}=0$ then we need to chose $c_{1d,r}=0.96$  in order to reproduce
the $\sigma$ width, i.e. $c_{1d,r}$ is a factor of $\sqrt{2}$
larger than that employed by Soto {\it et al.}, because our analytic expression for the width is a factor of two smaller. However, the introduction of a finite $f_{2p,r}$ ultimately permits a markedly smaller $c_{1d,r}$ to yield the
observed width.

\section{S-wave pion-pion scattering at leading order in $\chi$PT$_S$}\label{sec:pipi}

We now investigate the $\pi\pi\rightarrow\pi\pi$ scattering process in $\chi$PT$_S$
from threshold through the energies at which the $\sigma$-resonance affects
the phase shifts. 

\begin{figure}
\begin{center}
\subfigure{\includegraphics[width=0.5\textwidth,angle=0]{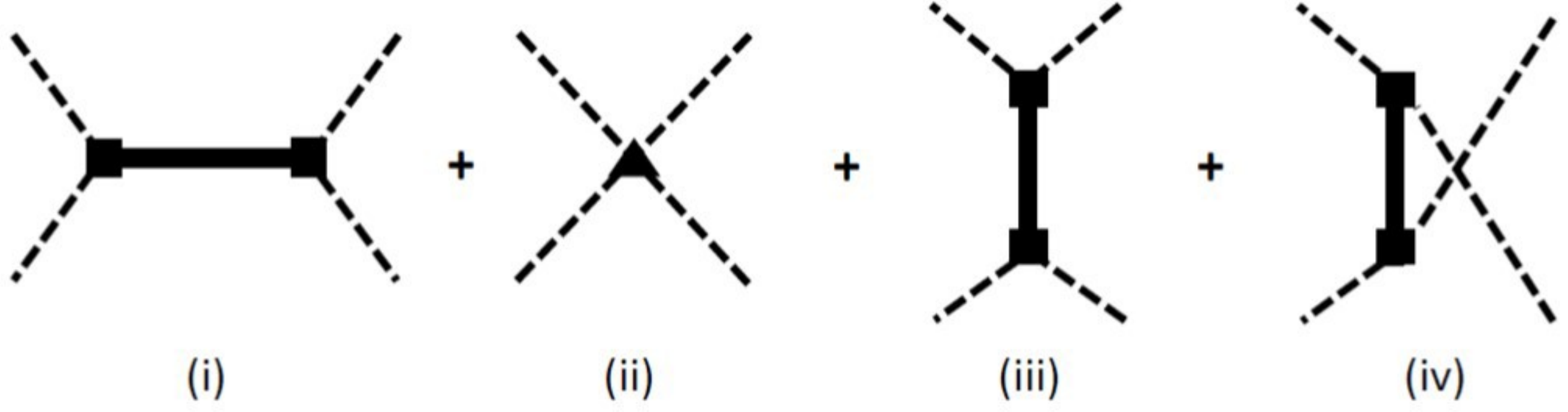}}
\end{center}
\caption{Tree-level diagrams contributing to $\pi\pi$ scattering in $\chi$PT$_S$:
(i) $s$-channel, (ii) contact term, 
(iii) $t$-channel, and (iv) $u$-channel. The triangle represents the interaction 
from the standard $\chi$PT $\cL_2$ Lagrangian.
The thick solid line indicates the dressed $\sigma$ propagator of Eq.~(\ref{alpha4}). The first two diagrams form 
the LO amplitude in our calculation, while the other two are part of the NLO amplitude.}
\label{ppfig2}
\end{figure}

Consider the diagrams (i)-(iv) of Fig.~\ref{ppfig2}. The thick line
indicates that we have resummed the $\sigma$ self energy and so are employing the propagator
(\ref{alpha4}) in all three diagrams. However, diagrams (iii) and (iv) are
formally next-to-leading order (NLO): the power counting  
assigns them an order $p^4/M_\sigma^2$, where $p \sim m_\pi$ in the threshold region and $p \sim M_\sigma$
in the resonance region. 

In contrast, the LO mechanisms are diagram (ii)---the tree-level $\chi$PT $\pi \pi$ scattering amplitude---near
threshold, where it is $\mathcal{O}(p^2)$, and diagram (i)---the $s$-channel $\sigma$ pole---in the resonance region, where it is
$\mathcal{O}(p^0)$. 
By combining diagrams (i) and (ii) we obtain an amplitude that is LO in both the threshold and resonance regions, and
interpolates smoothly between the two. 

The isospin $I$=0 projected pion-pion scattering amplitude at LO
is then
\beqa
\label{alpha7}
T^{I=0}(s,t,u)=\frac{1}{F^2} \bigg(3(s-m_\pi^2)+(t-m_\pi^2)+(u-m_\pi^2)-
\frac{12 c_{1d,r}^2 (s-2m_\pi^2)^2}
 {s-m_{S,r}^2-2 f_{2p,r} s^2-c_{1d,r}^2\tilde{\Sigma}(s)}\bigg).
 \eeqa 
 This amplitude is only perturbatively unitary: diagram (i) is unitary on its own, but no loop effects associated with diagram (ii) are included in our LO calculation, they enter only
 at $\mathcal{O}(p^4/\Lambda^2_{\chi {\rm SB}})$ in the chiral expansion, see also the discussion of the breakdown of this EFT below. Given this, we must use the first-order relation between the S-wave $\pi\pi$ phase shift $\delta^0_0$ and 
$T^{I=0}$~\cite{Taylor}:
\beqa
\delta^0_0=\frac{|{\bf k}|}{32\pi\sqrt{s}}\int_{-1}^1 d(\cos\theta)~\Re[T^{I=0}(s,\cos\theta)]~,
\label{eqint6}~
\eeqa
where $|{\bf k}|$=$\sqrt{s-4m_\pi^2}/2$ represents the magnitude of the center-of-mass (CM) momentum
and $\theta$ the CM scattering angle. The isoscalar $\pi \pi$ scattering length is then defined by:
\begin{equation}
a^0_0=\lim_{|{\bf k}| \rightarrow 0} \frac{\delta_0^0}{|{\bf k}|}.
\end{equation}
It is conventional to quote the $\pi \pi$ scattering lengths in units of $m_\pi^{-1}$.

\begin{figure}[h]
        \subfigure{
           \includegraphics[width=0.5\textwidth,angle=0]{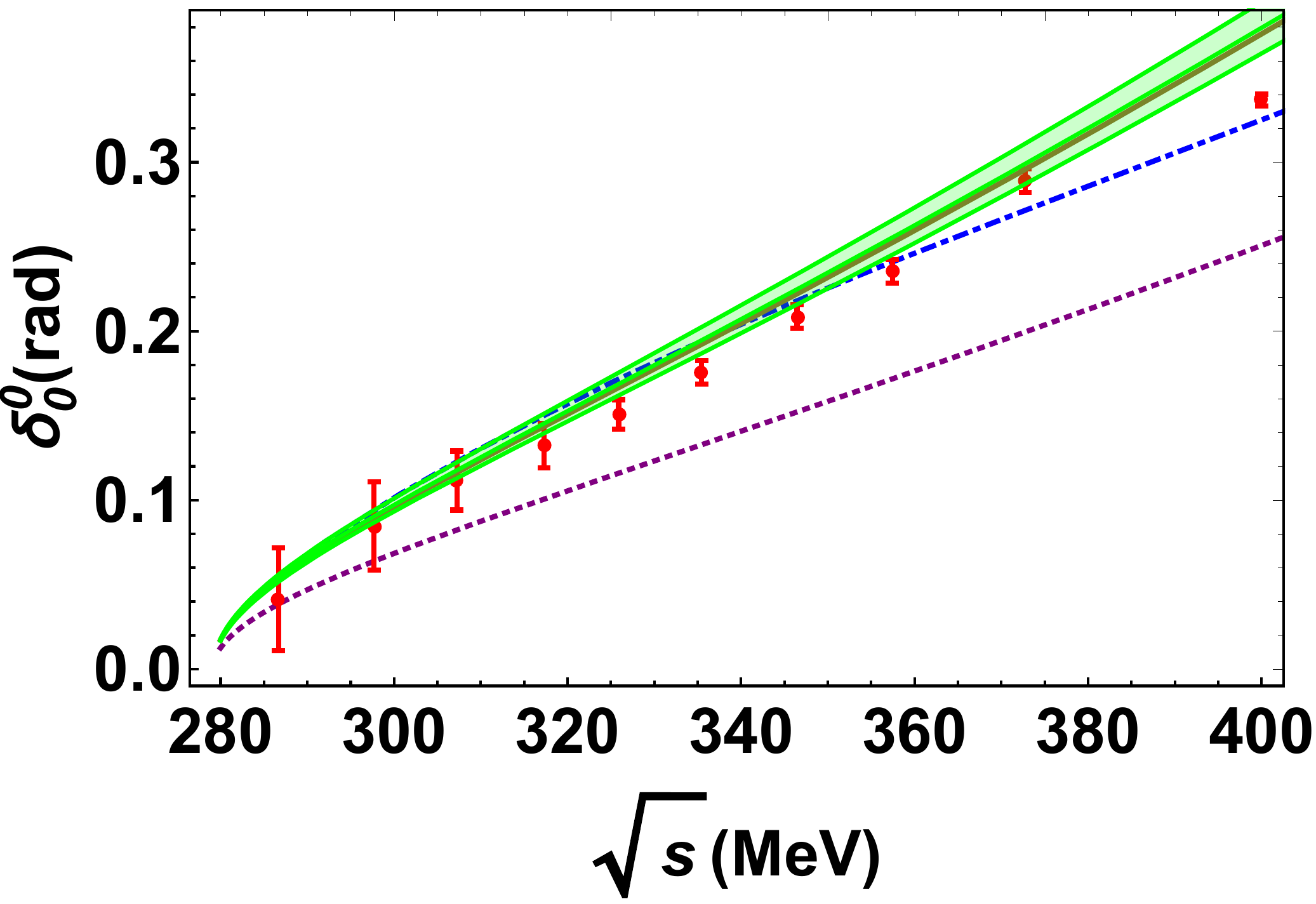}}
           \subfigure{\includegraphics[width=0.5\textwidth,angle=0]{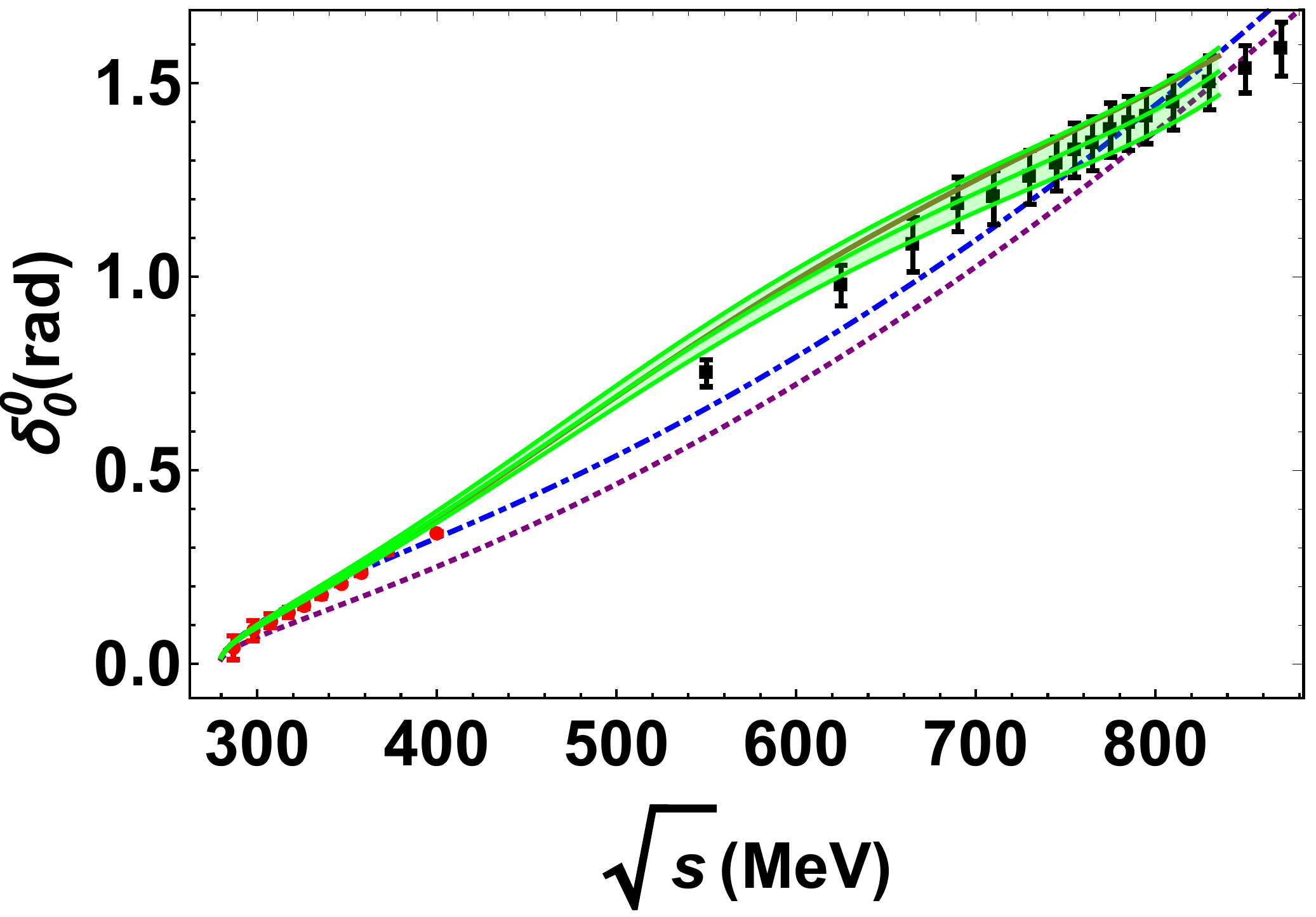}}
\caption{The $\pi\pi$ scattering phase shift as a function of the CM energy. In the left panel we show 
the results immediately above threshold, while the right panel shows the result up to $\sqrt{s}=870$ MeV. 
In both panels the dashed purple line represents the standard $\chi$PT result and
the dashed-dotted blue line is the combined result of diagrams (i) and (ii). This is to be compared to 
 the red circles (black squares) that represent the data from Ref.~\cite{Batley:2010} (Ref.~\cite{Protopopescu:1972sp}). The solid green and brown curves are results 
from the dispersive 
analyses of Refs.~\cite{Colangelo:2001df,GarciaMartin:2011cn} respectively. The green shaded band is a parameterization of the error reported in
Ref.~\cite{Colangelo:2001df}. \label{alphafig1}}
\end{figure}

Figure~\ref{alphafig1} shows the standard LO $\chi$PT [$\mathcal{O}(p^2)$]
result in the dashed purple curve and the total LO $\chi$PT$_S$ phase shift in the dashed-dotted
blue curve. We find that the 
contributions from the $\sigma$-meson physics are generally smaller than the $\mathcal{O}(p^2)$ $\chi$PT result. Thus, 
although the $s$-channel $\sigma$-meson pole exists, it 
only affects the total $\pi \pi$ phase shift weakly. 
In Fig.~\ref{alphafig1} we also compare our LO result to the dispersive/Roy-equation analyses from Refs.~\cite{Colangelo:2001df,GarciaMartin:2011cn} (solid green  and  brown curves). And we display $\pi \pi$ phase-shift data. The left panel emphasizes the lower-energy range, where data (red circles) were obtained by analyzing the $\pi\pi$ scattering in the final-state interactions between pions in the K$_{e4}$ decay
K$^{\pm}\rightarrow\pi^{\pm}\pi^{\mp}$e$^\pm\nu$~\cite{Batley:2010}. The description of these near-threshold data is very good---especially considering this is only a LO calculation. The addition 
of the $s$-channel $\sigma$-meson pole is enough to ameliorate the discrepancy between the $\mathcal{O}(p^2)$
$\chi$PT result and the data.

In the right panel we compare to data (black squares) in the energy range above 500 MeV from Ref.~\cite{Protopopescu:1972sp}, obtained from analysis of the reactions
$\pi^+p$~$\rightarrow\pi^+\pi^-\Delta^{++}$ and $\pi^+p$~$\rightarrow K^+K^-\Delta^{++}$. Adding the $s$-channel $\sigma$ brings the total phase shift
closer to these data, although there is somewhat of a difference in the curvature at higher energies between the data and the LO $\chi$PT$_S$ amplitude.
This difference is clear if one compares the dispersive results for $\delta_0^0$ to our calculation.

One might be concerned that the improved agreement in the threshold region comes at the cost of diminished performance for the $I=2$ $\pi \pi$ scattering length $a_{\pi\pi}^{I=2}$, where $\chi$PT's tree-level prediction is in remarkable agreement with the experimental data. However, since the $\sigma$-meson propagator only enters the LO amplitude in the $s$-channel it actually has no impact on the $I=2$ phase shift, and the $\chi$PT LO result for $a_0^2$ is preserved in this LO $\chi$PT$_S$ calculation. The $t-$ and $u$-channel $\sigma$-meson poles are part of the NLO $\chi$PT$_S$ $\pi \pi$ amplitude. Together with other NLO effects they will produce a small shift in the constants (\ref{eq:parameters}), as already discussed in general terms in Sec.~\ref{sec:sigmaSigma}. 

We now discuss higher-order effects like these graphs. We will see that some NLO pieces of the amplitude have particular impact at the higher energies shown in the right panel of Fig.~\ref{alphafig1}.
Unlike the standard $\chi$PT amplitude, Eq.~(\ref{alpha7}) 
does not respect crossing symmetry---even before the isospin projection is made. 
The NLO graphs (iii) and (iv), with a dressed $\sigma$ propagator, restore crossing symmetry. The
additional amplitude in the $I=0$ channel is:
\begin{equation}
\Delta T^{I=0}(s,t,u)=-\frac{4~c_{1d,r}^2}{F^2} \left(
\frac{(t-2m_\pi^2)^2}
 {t-m_{S,r}^2-2 f_{2p,r} t^2-c_{1d,r}^2\tilde{\Sigma}(t)}
 + \frac{(u-2m_\pi^2)^2}
 {u-m_{S,r}^2-2 f_{2p,r} u^2-c_{1d,r}^2\tilde{\Sigma}(u)}\right).
 \label{alpha7b}
 \end{equation}
 
  \begin{figure}[h]
     \centerline{   
           \includegraphics[width=0.60\textwidth,angle=0]{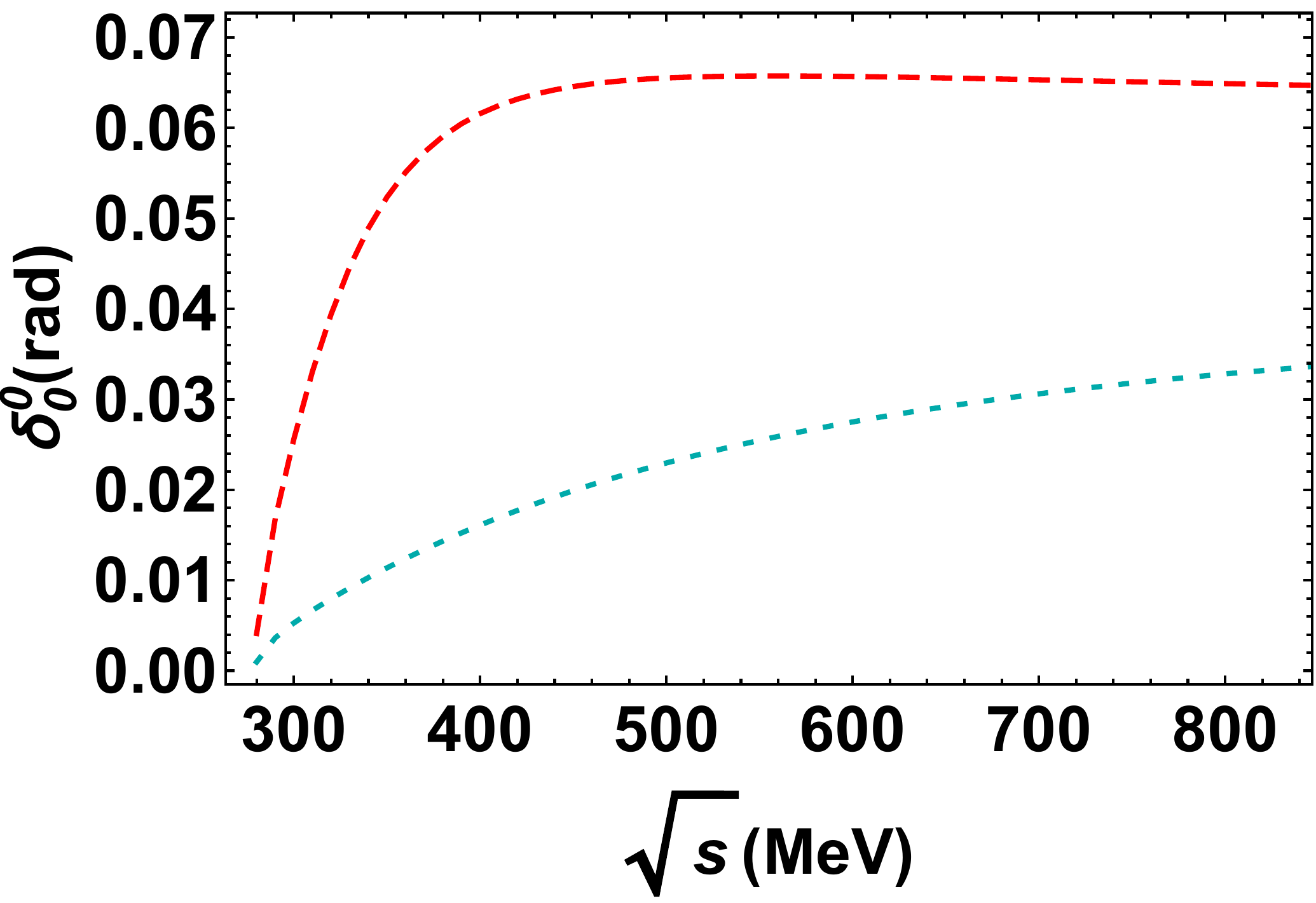}}
\caption{The $\pi\pi$ scattering phase shift due to graphs involving $\sigma$-meson exchange at tree level. The dashed red (dotted
cyan) line represents the $s$- ($t$- and $u$-) channel contribution. }
\label{fig:stu}
\end{figure}
\noindent
In Fig.~\ref{fig:stu}, we plot the $\pi\pi$ scattering phase shift as predicted by the $\sigma$-meson
part of the
amplitude of Eqs.~(\ref{alpha7}) and (\ref{alpha7b}). The plot shows the $s$-channel $\sigma$ contribution, the last term in Eq.~(\ref{alpha7}), as the dashed red curve,
and the combined $t$- and $u$-channel contributions, Eq.~(\ref{alpha7b}) as the dotted cyan curve.
Both these curves level
off as a function of the CM energy---due to the $1/q^4$ behavior of the $\sigma$ propagator at large momenta. This calculation shows that the $t$- and $u$-channel $\sigma$-pole have a markedly smaller effect on the $I=0$ S-wave phase shift than does the $s$-channel $\sigma$-pole. The relative size of these effects is consistent with our assignment of these graphs to the NLO piece of the $\chi$PT$_S$ amplitude. We therefore sacrifice crossing symmetry in order to have our EFT encode the hierarchy of $\sigma$-meson mechanisms for $s$ between $4 m_\pi^2$ and $\sim M_\sigma^2$.

As already observed, our amplitude violates unitarity. Since the standard $\chi$PT $\mathcal{O}(p^2)$ amplitude is the largest piece of the S-wave phase shift it drives this violation. It violates the simplest consequence of unitarity already for $\sqrt{s}$ slightly below $700$ MeV~\cite{DonoghueGolowichHolstein}. Of course, unitarity is restored order-by-order in the $\chi$PT expansion, so these defects are somewhat remedied by loop graphs at $\mathcal{O}(p^4)$, but those are not included here. This calculation is thus certainly limited in scope to $\sqrt{s} < 700$ MeV, even though we show a wider range here. At higher orders in the theory it may be possible to describe data all the way up to $\sqrt{s}$ of order the rho-meson mass. However, the size of the phase shift for $\sqrt{s} \geq 600$ MeV implies that the ${\mathcal O}(p^4)$ $\chi$PT amplitude will already produce marked corrections to the LO result in that region, so the LO calculation we have done here cannot be trusted beyond $\sqrt{s}=600$ MeV. 

Finally, we comment on the role of the $\rho(770)$ in our approach. 
We have elevated the $\sigma$ to the status of a dynamical field, but continued to integrate the $\rho$ out and incorporate its effects through $\mathcal{O}(p^4)$ contact interactions. Those effects could, in principle, affect the $I=0$ and $I=2$ scattering lengths at NLO through the combination of LECs $\bar{l}_1 + 2 \bar{l}_2$. However, in the resonance-saturation approach of Ref.~\cite{Ecker:1988te} the tree-level contribution of the $\rho$ to $\bar{l}_1 + 2 \bar{l}_2$ equals zero. 

\section{$\gamma\gamma\rightarrow\pi^0\pi^0$ scattering cross section in $\chi$PT$_S$}

\label{sec:gammagammapipi}

\begin{figure}[h]
\centering
\includegraphics[scale=0.3]{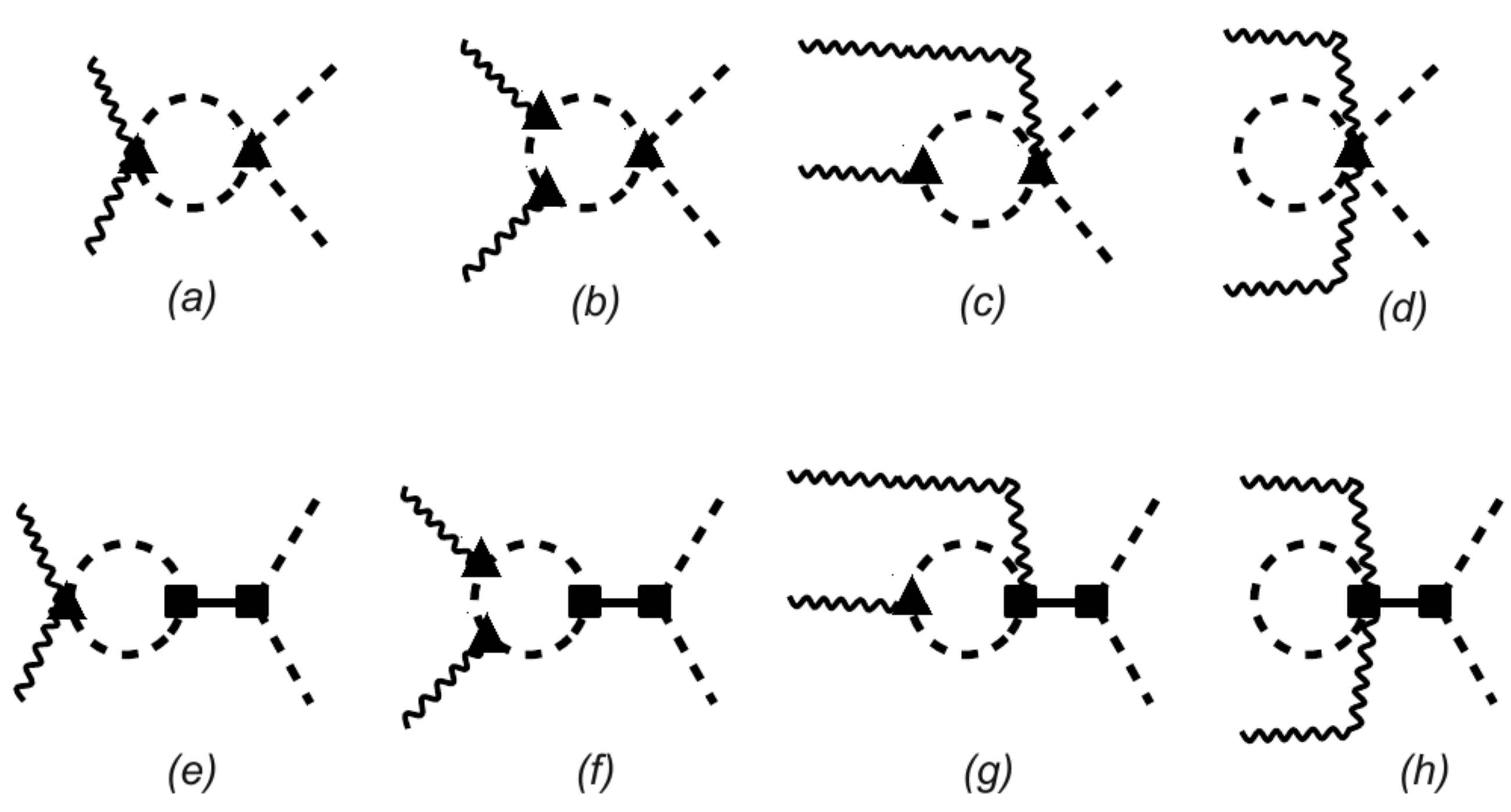}
\caption{The leading-order diagrams for the process $\gamma\gamma\rightarrow\pi^0\pi^0$ in $\chi$PT$_S$.
The upper four correspond to standard $\chi$PT while the lower four appear additionally in $\chi$PT$_S$.  The wavy lines are photons.
The direction of time is to the right.\label{ppfig4}}
\end{figure}

There are no $\mathcal{O}(p^2)$ (tree-level) contributions to the reaction $\gamma \gamma \rightarrow \pi^0 \pi^0$. Note also that the $\sigma$ 
is not charged, so minimal substitution does not generate any tree-level couplings between it and photons. This process therefore must involve pion-loop contributions, and in $\chi$PT$_S$ these come 
in two varieties: diagrams with a $\sigma$ pole and diagrams without such a pole. 

The top line of Fig.~\ref{ppfig4} shows the  $\mathcal{O}(p^4)$ contributions to the process $\gamma\gamma\rightarrow\pi^0\pi^0$ of the first type. These are the standard 
$\chi$PT graphs at this order. In $\chi$PT$_S$ the bottom four graphs---again with a dressed $\sigma$ propagator---are part of the LO amplitude if we consider the region
$s \sim M_\sigma^2$. 

\begin{figure}[h]
\centering
\includegraphics[scale=0.3]{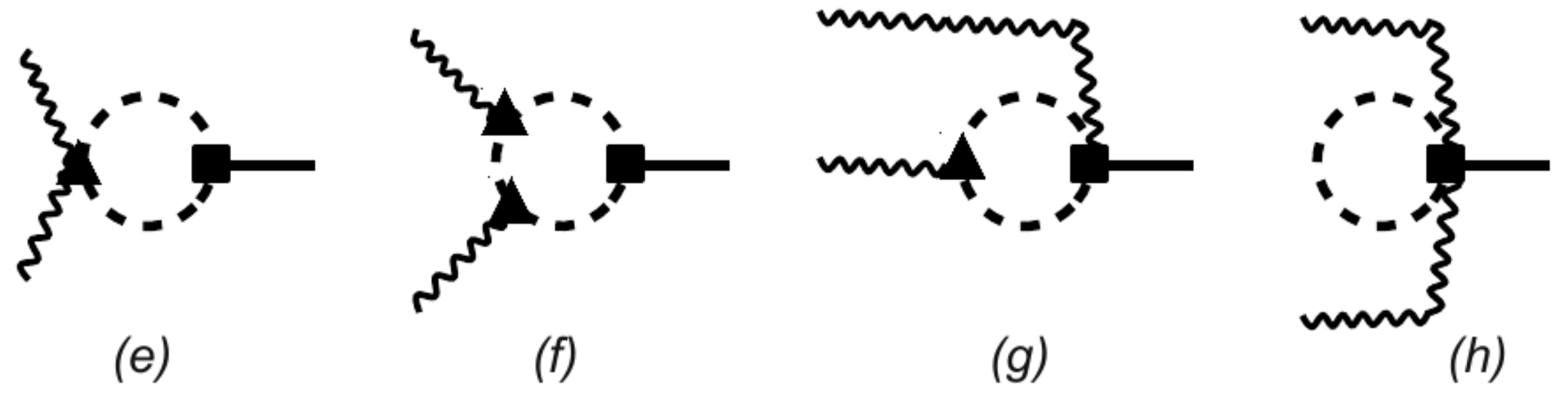}
\caption{The diagrams that are represented by i$\mathcal{V}$ in Eq.~(\ref{Ceq196}).  \label{Cfig7}}
\end{figure}

Before calculating the amplitude for this process we first verify the Ward identity. This states that 
the scattering amplitude $\epsilon_\mu \mathcal{V}^\mu$ for a
process that includes an external photon with a polarization $\epsilon_\mu$ and momentum $k^\mu$
gives zero when the polarization vector of the photon is replaced by its momentum, i.e., $k_\mu\mathcal{V}^\mu$=0.
This holds true for any number of external photons
when the associated polarization vectors are replaced by the corresponding momenta. The Ward identity
for the upper four standard $\chi$PT diagrams of Fig.~\ref{ppfig4}
has been verified in Ref.~\cite{Donoghue:1988eea}. Here we derive the Ward identity for the bottom four graphs. 
The general form of the amplitude for these graphs can be written as
\beqa
\label{Ceq196}
i\mathcal{T}=i\mathcal{V}\bigg(\frac{i}{s-m_{S,r}^2-2 f_{2p,r} s^2-c_{1d,r}^2\tilde{\Sigma}(s)}\bigg)\bigg(\frac{i2c_{1d,r}(s-2m_\pi^2)}{F}\bigg) ~,
\eeqa
where $\mathcal{V}$=$\mathcal{V}_e$+$\mathcal{V}_f$+$\mathcal{V}_g$+$\mathcal{V}_h$ represents the  $\sigma$-irreducible-vertex for $\gamma \gamma \rightarrow \sigma$ graphs shown in  Fig.~\ref{Cfig7} and the subscripts on each $\mathcal{V}$ indicate the corresponding diagram. If the vertex $\mathcal{V}$ obeys the Ward identity then the entire amplitude will obey it. 

Suppose $k_{1\mu}$
and
$k_{2\mu}$ ($\epsilon_{1\mu}$ and $\epsilon_{1\mu}$) represent the four momenta 
(polarization vectors) of the external photons
of the graphs in Fig.~\ref{Cfig7} such that $s$=$(k_{1\mu}+k_{2\mu})^2$.
We can verify the Ward identity by
replacing either $\epsilon_{1\mu}$ or 
$\epsilon_{2\mu}$. Here we do it for $\epsilon_{1\mu}$; the result for
$\epsilon_{2\mu}$ follows from (1$\longleftrightarrow$2) symmetry. Once the replacement 
$\epsilon_{1\mu}\rightarrow k_{1\mu}$ has been made, the amplitudes $\mathcal{V}_e$--$\mathcal{V}_h$ are no longer the same. 
We will call the results of the replacement the \textit{transformed} amplitudes and label them with a superscript 
\textit{W}, i.e., $\mathcal{V}^W_e$--$\mathcal{V}^W_h$.
 
The sum of the transformed amplitudes corresponding to Fig.~\ref{Cfig7}(e)
and Fig.~\ref{Cfig7}(f) can be written as:
\beqa
\label{Ceq176}
i\mathcal{V}_e^W+i\mathcal{V}_f^W&=&\frac{8c_{1d,r}e^2}{F}\bigg(-\frac{2}{d}\bigg)\epsilon_{2\nu}k_1^\nu\int\frac{d^4q}{(2\pi)^4}
\frac{q^2}{[q^2-m_\pi^2]^{2}}~,
\eeqa
where $d$ represents the dimension. Similarly, the sum of the transformed amplitudes corresponding to Fig.~\ref{Cfig7}(g)
and Fig.~\ref{Cfig7}(h) can be written as:
\beqa
\label{Ceq180}
i\mathcal{V}_g^W+i\mathcal{V}_h^W&=&\frac{8c_{1d,r}e^2}{F}\epsilon_{2\nu}k_1^\nu\int\frac{d^4q}{(2\pi)^4}
\frac{1}{q^2-m_\pi^2}~.
\eeqa
Adding up all the transformed amplitudes, employing dimensional regularization, 
and using $\frac{1}{d}\simeq\frac{1}{4}(1+\frac{\epsilon}{4})$, we get, as $\epsilon\rightarrow0$
\beqa
\label{Ceq195}
i\mathcal{V}^W&=&\frac{8c_{1d,r}e^2}{F}\bigg(-\frac{1}{2}\bigg)\epsilon_{2\nu}k_1^\nu\frac{i(-m_\pi^2)}{(4\pi)^2}
\bigg[-\frac{4}{\epsilon}-2+2\gamma-2\ln \bigg(\frac{4\pi
\mu^2}{m_\pi^2}\bigg)\bigg]\no\\*
&&+\frac{8c_{1d,r}e^2}{F}\epsilon_{2\nu}k_1^\nu\frac{i(-m_\pi^2)}{(4\pi)^2}\bigg[-\frac{2}{\epsilon}-1+\gamma-\ln
 \bigg(\frac{4\pi\mu^2}{m_\pi^2}\bigg)\bigg]\no\\*
  &=&0~.
\eeqa
This verifies the Ward identity.

Turning now to the $\gamma\gamma\rightarrow\pi^0\pi^0$ cross section, the amplitude for the top
four standard $\chi$PT
graphs of Fig.~\ref{ppfig4} is evaluated in Ref.~\cite{Donoghue:1988eea} and here we simply recycle their results for that part
of the amplitude. 
Then, the differential scattering cross section for the $\gamma\gamma\rightarrow\pi^0\pi^0$ process
in $\chi$PT$_S$ can be expressed in the CM frame as
\beqa
\frac{d\sigma}{d\Omega}=\frac{1}{64\pi^2s}\sqrt{1-\frac{4m_\pi^2}{s}}\langle|\mathcal{T}|^2\rangle~,
\label{eq201}
\eeqa
where 
\beqa
\label{eq200}
\langle|\mathcal{T}|^2\rangle=\frac{1}{4}\Bigg(\frac{s^2}{2}+sm_\pi^2+m_\pi^4\Bigg)\Big(|H_{\chi PT}(s)+H_\sigma(s)|^2\Big) ~.
\eeqa
The standard $\chi$PT and additional part of the amplitude that arise in $\chi$PT$_S$ ($H_\sigma(s)$) are:
\beqa
H_{\chi PT}(s)=-\frac{1}{8\pi^2}\frac{2e^2}{F^2}\frac{s-m_\pi^2}{s}
\Bigg\{1+\frac{m_\pi^2}{s}\Bigg[\ln\Bigg(\frac{x_+}{x_-}\Bigg)-i\pi\Bigg]^2\Bigg\}~,
\label{196a}
\eeqa
and 
\beqa
H_\sigma(s)&=&\frac{2c_{1d,r}^2e^2}{F^2\pi^2}~
\Bigg[\frac{1}{4s^2}
\Bigg\{-2m_\pi^2(s+2m_\pi^2\log(m_\pi^2))+\frac{1}{3}\Bigg(18m_\pi^2s-2s^2+12m_\pi^4\log(m_\pi^2)\no\\*
&&~~~~+12m_\pi^2(s-2m_\pi^2)\bigg[Li_2\bigg(\frac{1}{x_+}\bigg)+
Li_2\bigg(\frac{1}{x_-}\bigg)\bigg]\Bigg)\Bigg\}-\frac{1}{3}\Bigg]
\bigg(\frac{(s-2m_\pi^2)}{s-m_{S,r}^2-2 f_{2p,r}s^2-c_{1d,r}^2\tilde{\Sigma}(s)}\bigg)~,\label{alpha8}\no\\*
\eeqa
with $x_{\pm}$ given by
\beqa
x_\pm&=&\frac{1}{2}\pm\frac{1}{2}\sqrt{1-\frac{4m_\pi^2}{s}}.
\label{eq134f}
\eeqa
and $Li_2$ representing the dilogarithm function.

\begin{figure}[h!]
\centering
\includegraphics[scale=0.385]{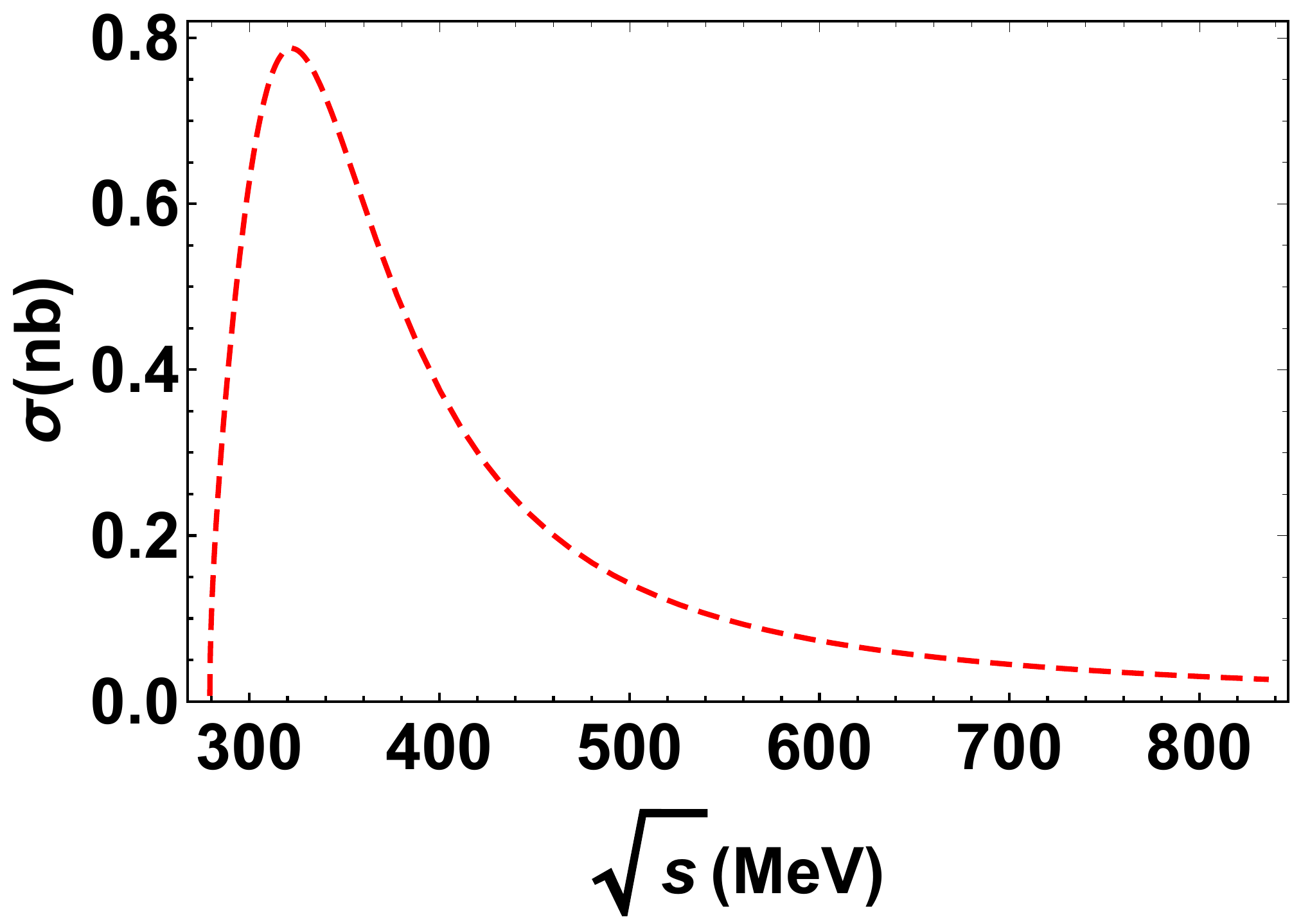}
\includegraphics[scale=0.385]{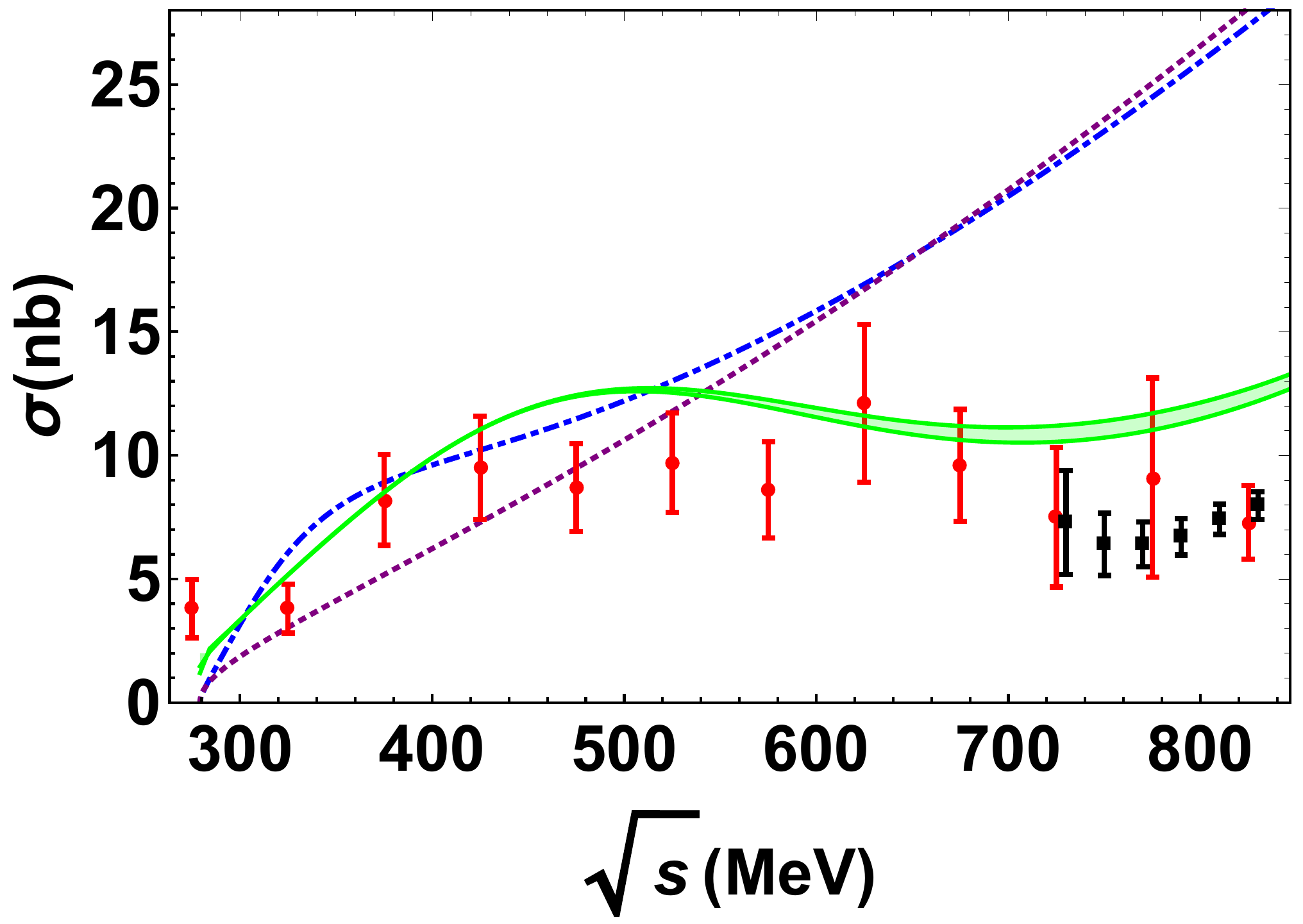}
\caption{The $\gamma\gamma\rightarrow\pi^0\pi^0$ cross section as a function of the CM energy for $|\cos\theta| \leq 0.8$, where $\theta$ is the CM scattering angle.
The dashed red curve in the left panel is the contribution 
from only the bottom graphs of Fig.~\ref{ppfig4}. Meanwhile the dotted purple curve in the right panel is the leading contribution in $\chi$PT, i.e., the top graphs in
Fig.~\ref{ppfig4}.
The dashed-dotted blue curve in the right panel is the total
combined result at LO in $\chi$PT$_S$.
Experimental data are from Ref.~\cite{Marsiske:1990hx} (red circles) and Ref.~\cite{Uehara:2008ep} (black squares).
The green band represents the once-subtracted result obtained in Ref.~\cite{Hoferichter:2011wk} from 
a dispersive Roy-equation analysis.\label{alphafig2}}
\end{figure}

In Fig.~\ref{alphafig2}, the left graph shows the $\gamma\gamma\rightarrow\pi^0\pi^0$
cross section due to the lower four graphs of Fig.~\ref{ppfig4}, those that involve the $\sigma$-meson pole. 
The bump seen there is inherited by the result for the total cross section represented by
the dashed-dotted blue curve in the right panel, which has some signal of the $\sigma$ resonance near 400 MeV.
This signal produces a good match between the LO $\chi$PT$_S$ $\gamma \gamma \rightarrow \pi^0 \pi^0$ cross section
and that obtained in a Roy-equation treatment of this reaction (with one subtraction)~\cite{Hoferichter:2011wk} up to $\sqrt{s} \approx 550$ MeV. The latter is represented
in Fig.~\ref{alphafig2} by the green band. 

The cross-section data for  $\gamma\gamma\rightarrow\pi^0\pi^0$ process
have been measured by The Crystal Ball Collaboration and reported in Ref.~\cite{Marsiske:1990hx}.
They analyzed the $e^+e^-\rightarrow e^+e^-\gamma^*\gamma^*\rightarrow e^+e^-\pi^0\pi^0$
reaction from threshold to about 2 GeV to obtain the cross section. The paper 
reports that the pion detection efficiency drops considerably for $|\cos\theta|>0.8$, and therefore
they have restricted their extraction of the $\gamma \gamma \rightarrow \pi^0 \pi^0$ cross section to the region $|\cos\theta|\leq0.8$.
Since the differential cross section in Eq.~(\ref{eq201}) is independent of
the scattering angle, the total cross section in the region 
$|\cos\theta|\leq0.8$ for the $\gamma\gamma\rightarrow\pi^0\pi^0$ process can be written as 
\beqa
\sigma&=&\frac{1}{2}3.2\pi\frac{d\sigma}{d\Omega},\no
\label{eq230}
\eeqa
where the factor of $1/2$ accounts for the identicality of the final-state particles.
 The enhancement that we attribute to the $s$-channel $\sigma$-meson pole 
is  slightly visible in the data (red circles) from Ref.~\cite{Marsiske:1990hx}. Our result for the total cross section agrees with this data to within 1.5 standard deviations up to $\sqrt{s} \approx 550$ MeV. 

At higher energies our LO result and the Roy-equation result of Ref.~\cite{Hoferichter:2011wk} have very different energy dependence. 
As already discussed in Sec.~\ref{sec:pipi}, the absence of $\pi \pi$ loop graphs means our LO amplitude is not correct once the $\pi \pi$ phase shift becomes significant, and the energy dependence obtained at LO in this theory is not a good match for the Roy-equation parameterization once $\sqrt{s} \geq 600$ MeV. This also means we cannot describe the
higher-statistics, higher-energy data on $\gamma \gamma \rightarrow \pi^0 \pi^0$ obtained in  Ref.~\cite{Uehara:2008ep}. Those data are represented by the black squares in Fig.~\ref{alphafig2}. 

Refs.~\cite{Ametller:2014vba,Ametller:2015xva}  obtained good agreement with the $\gamma \gamma \rightarrow \pi^0 \pi^0$ and $\gamma \gamma \rightarrow \pi^+ \pi^-$ data, respectively. However, they achieved this 
by adopting a $\sigma$-meson propagator
\beq
D(s)=\frac{1}{s - M_\sigma^2 + i \Gamma(s) M_\sigma}, 
\eeq
with $\Gamma(s)=\left(\frac{s-s_0}{M_\sigma^2 - s_0}\right)^{1/2} \Gamma_0$ the energy-dependent width of the $\sigma$~\cite{Ametller:2015xva}. They 
then adjusted the $\sigma \pi \pi$ coupling that enters the numerator of the $\sigma$-pole diagram in $\gamma \gamma \rightarrow \pi \pi$.
But unitarity requires that the $\sigma \pi \pi$ coupling is the mechanism by which the width $\Gamma$ is generated. Ametller and Talavera's
approach to the two $\gamma \gamma \rightarrow \pi \pi$ reactions therefore corresponds to a $\pi \pi$ amplitude that is not unitary. 

Finally, we examine the pion electromagnetic polarizabilities. The stiffness of a pion against
deformation by external electromagnetic fields is characterized by dipole and
quadrupole polarizabilities. The Compton-scattering reaction $\gamma\pi\rightarrow\gamma\pi$ seems the obvious process from which to extract these quantities,
but they can be obtained from $\gamma\gamma\rightarrow\pi\pi$ as well, since the two reactions
are related by crossing symmetry. The $(\alpha_1-\beta_1)_{\pi^0}$ difference of dipole and
$(\alpha_2-\beta_2)_{\pi^0}$ difference of quadrupole polarizabilities are defined \cite{Donoghue:1988eea}
through the expansion of the
amplitude $H(s)$=$H_{\chi PT}(s)$+$H_\sigma(s)$ about $s$=0 as
\begin{equation}
\frac{1}{4\pi m_\pi}H(s)=(\alpha_1-\beta_1)_{\pi^0}+\frac{s}{12}(\alpha_2-\beta_2)_{\pi^0}
+\mathcal{O}(s^2)~.
\label{NEWeq231}
\end{equation}
Our results for these polarizabilities are
\begin{eqnarray}
(\alpha_1-\beta_1)_{\pi^0}&=&- \frac{e^2}{192 F^2 \pi^3 m_\pi} - \frac{e^2 c_{1d,r}^2 m_\pi}
{12 F^2 \pi^3 (m_{S,r}^2 + c_{1d,r}^2 \tilde{\Sigma}(0))},\\
(\alpha_2-\beta_2)_{\pi^0}&=&\frac{13 e^2}{240F^2\pi^3m_\pi^3}
+\frac{13 e^2 c_{1d,r}^2}{15 F^2 \pi^3 m_\pi (m_{S,r}^2 + c_{1d,r}^2 \tilde{\Sigma}(0))}
-\frac{e^2 c_{1d,r}^2 m_\pi (1-c_{1d,r}^2 \tilde{\Sigma}'(0))}
{F^2 \pi^3  (m_{S,r}^2 + c_{1d,r}^2 \tilde{\Sigma}(0))^2},
\end{eqnarray}
where $\tilde{\Sigma}'(0)=\left. \frac{d \tilde{\Sigma}(s)}{d s}\right|_{s=0}$ is dimensionless.
Note that the LEC $f_{2p,r}$ does not appear here, since it only gives the $s^2$ dependence of the inverse $\sigma$-meson propagator. Its value does, however, 
affect the polarizabilities, since changes in $f_{2p,r}$ result in changes in $c_{1d,r}$ and $m_{S,r}$ so that the renormalization conditions are maintained. 

\begin{table}[h]
\caption{The dipole and quadrupole polarizabilities in units of $10^{-4}$ fm$^3$ and $10^{-4}$ fm$^5$.
The second and third columns contain the standard $\chi$PT one-loop and two-loop results
from Ref.~\cite{Donoghue:1988eea} and Ref.~\cite{Gasser:2005ud} respectively. The fourth column contains the results
from dispersion-relation calculations~\cite{Gasser:2005ud} The last column is
our $\chi$PT$_S$ result at one-loop. }
\vspace{3mm}
\centering
\begin{tabular}{|c|c|c| c|c| }
\hline
Polarizabilities& $\chi$PT to &$\chi$PT to&Disperson&$\chi$PT$_S$ at\\
&  one-loop &two-loop &relation& one-loop\\
\cline{1-5}
$(\alpha_1-\beta_1)_{\pi^0}$&-0.98\cite{Donoghue:1988eea} &-1.9\cite{Gasser:2005ud}&-1.6\cite{Gasser:2005ud}&-1.1\\
\cline{1-5}  
$(\alpha_2-\beta_2)_{\pi^0}$&20.37\cite{Donoghue:1988eea} &37.6\cite{Gasser:2005ud}&39.7\cite{Gasser:2005ud}&21.6\\
\cline{1-5}  
\hline
\end{tabular}
\label{tab2}
\end{table}

Using our values for $c_{1d,r}$ and $m_{S,r}$ we obtain the dipole and quadrupole polarizabilites 
presented in Table~\ref{tab2}. For comparison purposes, we have also presented
the values from standard $\chi$PT one-loop [$\mathcal{O}(p^4)$], two-loop [$\mathcal{O}(p^6)$], and dispersion-relation calculations.
We see from Table~\ref{tab2} that our calculation does capture some physics beyond the standard one-loop calculation, and seems
to incorporate some of the two-loop physics that gives large corrections to both
$(\alpha_1 - \beta_1)_{\pi^0}$ and $(\alpha_2 - \beta_2)_{\pi^0}$. However, a
calculation with $f_{2p,r}=0$ (and $c_{1d,r}$ re-adjusted to again reproduce the $\sigma$ width) bridges half the gap between the $\mathcal{O}(p^4)$ and $\mathcal{O}(p^6)$
polarizabilities. Of course, this is at the cost of an unphysically large $\pi \pi$ and $\gamma \gamma \rightarrow \pi \pi$ cross section.

\section{Conclusion}

\label{sec:conclusion}

In this paper we have shown that an EFT in which standard $\chi$PT is augmented by the addition of a light scalar field, worked out initially by Soto, Talavera, and Tarr\'us in Ref.~\cite{Soto:2011ap}, provides a 
consistent and accurate leading-order description of the $\sigma$-meson pole, the isoscalar $\pi \pi$ scattering length, and the data for $\pi \pi$ scattering and $\gamma \gamma \rightarrow \pi^0 \pi^0$ 
up to center-of-mass energies $\approx 500$ MeV. This obviates the need for the inconsistent treatment of the $\pi \pi$ amplitude in the latter reaction that was adopted in Ref.~\cite{Ametller:2014vba}.
We also found that the analytic result of Refs.~\cite{Soto:2011ap,Hansen:2016fri} for the $\sigma$-meson width is too large by a factor of two. 

We use a Dyson equation to resum the $\chi$PT$_S$ self-energy correction to the
scalar-meson propagator in the vicinity of the resonance. Hence our approach generates a $\pi \pi$ amplitude in the scalar channel that is quite similar to that obtained in the inverse-amplitude method (IAM)~\cite{GomezNicola:2007qj,Hanhart:2008mx,Doring:2016bdr}. The IAM constructs a unitary $\pi \pi$ amplitude which reproduces both the LO and NLO $\chi$PT results, and (after suitable modification) respects the Adler zero in this channel.
However, the IAM does not include a modification to the $\sigma$ propagator that has it behaving as $s^{-2}$ for $s$ far from the $\sigma$ pole. The EFT treatment we have adopted here shows that such behavior is, in fact, mandated by the divergence structure of the diagram that generates the leading contribution to the $\sigma$'s width in the EFT. The fact that the $\sigma$ propagator has this unusual off-shell dependence in turn allows the $\sigma$ to be quite a weak effect once one considers the real energies that are a significant distance from the pole.

An interesting subject for future study would be to explicitly compare the $\pi \pi$ and $\gamma \gamma \rightarrow \pi \pi$ amplitudes obtained in this work and in studies using Roy equations~\cite{Ananthanarayan:2000ht,Hoferichter:2011wk} and the IAM~\cite{GomezNicola:2007qj}. By examining how these amplitudes behave as a function of $s$ as one moves from the $\sigma$-meson pole to the real axis where scattering is computed, and then to the sub-threshold region, one could determine the extent to which the simpler amplitude computed here reproduces the features obtained in these more sophisticated approaches. Our leading-order $\chi$PT$_S$ amplitude could also be compared to the results of Refs.~\cite{Achasov:2011,Achasov:2012}, wherein a phenomenological $\pi\pi$ scattering amplitude with good analyticity properties in the $s$-plane that matches the Roy-equation solution quite well was obtained from the linear $\sigma$ model and a simple background amplitude. 

Lastly, we observe that the power counting in which our leading-order calculation was derived has some issues if its accuracy is reviewed {\it a posteriori}. For example, the $s$-channel $\sigma$ pole is nominally the LO mechanism [$\mathcal{O}(p^0)$] for $s \sim M_\sigma^2$. However, the results for $\pi \pi$ scattering show that---after all the parameters are chosen---that $s$-channel $\sigma$ pole is a fairly small correction to the $\mathcal{O}(p^2)$ $\chi$PT amplitude in this region. Of course, this is because the $\sigma$-meson pole has moved so far from the real axis upon the inclusion of the the one-loop self energy. However, that significant movement itself raises  concerns, since the power counting employed here is for a narrow resonance, where $i \Gamma/2$ is a perturbative correction to the tree-level mass. It is not clear if the physical $\sigma$ meson satisfies this criterion.  Comparison of the EFT amplitude as a function of $s$ with that found in other approaches will help us understand this issue, since it will illuminate the extent to which the $s$-dependence of the amplitude arises from the one-loop self-energy effect we have focused on here. 

\section*{Acknowledgments} We thank Martin Hoferichter, Joan Soto, Carlos Schat, Matthias Schindler, and Pedro Talavera for valuable discussions. We are also grateful to Joan Soto for useful comments on the manuscript. This work was supported by the US Department of 
Energy under grant number DE-FG02-93ER-40756.

   \bibliographystyle{unsrt}

\begin{thebibliography}{9}
   \bibitem{Pelaez:2015qba} 
  J.~R.~Pel\'aez,
  Phys.\ Rept.\  {\bf 658}, 1 (2016)
  doi:10.1016/j.physrep.2016.09.001
  [arXiv:1510.00653 [hep-ph]].
  
     \bibitem{johnson} 
 M. H. Johnson and E. Teller, 
 Phys.\ Rev. {\bf 98}, 783 (1955).
 
 \bibitem{schwinger}
  J. S. Schwinger, 
  Annals \ Phys. {\bf 2}, 407 (1957).  
  
\bibitem{Agashe:2014kda} 
  K.~A.~Olive {\it et al.} [Particle Data Group Collaboration],
  Chin.\ Phys.\ C {\bf 38}, 090001 (2014).
 
 
  
\bibitem{Caprini:2005zr} 
  I.~Caprini, G.~Colangelo, and H.~Leutwyler,
  Phys.\ Rev.\ Lett.\  {\bf 96}, 132001 (2006)
  [hep-ph/0512364].
  
 
\bibitem{Colangelo:2001df} 
  G.~Colangelo, J.~Gasser, and H.~Leutwyler,
  Nucl.\ Phys.\ B {\bf 603}, 125 (2001)
  [hep-ph/0103088].
  
\bibitem{GarciaMartin:2011jx} 
  R.~Garcia-Martin, R.~Kaminski, J.~R.~Pel\'aez, and J.~Ruiz de Elvira,
  Phys.\ Rev.\ Lett.\  {\bf 107}, 072001 (2011)
  [arXiv:1107.1635 [hep-ph]].
  
\bibitem{Moussallam:2011zg} 
  B.~Moussallam,
  Eur.\ Phys.\ J.\ C {\bf 71}, 1814 (2011)
  [arXiv:1110.6074 [hep-ph]].
  

    \bibitem{Briceno:2016mjc} 
  R.~A.~Briceno, J.~J.~Dudek, R.~G.~Edwards and D.~J.~Wilson,
  Phys.\ Rev.\ Lett.\  {\bf 118}, no. 2, 022002 (2017)
  doi:10.1103/PhysRevLett.118.022002
  [arXiv:1607.05900 [hep-ph]].
  
 \bibitem{Crewther:2015dpa} 
  R.~J.~Crewther and L.~C.~Tunstall,
  PoS CD {\bf 15}, 132 (2015)
  [arXiv:1510.01322 [hep-ph]].
  
  
 \bibitem{Aoki:2014oha} 
  Y.~Aoki {\it et al.} [LatKMI Collaboration],
  Phys.\ Rev.\ D {\bf 89}, 111502 (2014)
  doi:10.1103/PhysRevD.89.111502
  [arXiv:1403.5000 [hep-lat]].
  
  \bibitem{Appelquist:2016viq} 
  T.~Appelquist {\it et al.},
  Phys.\ Rev.\ D {\bf 93}, no. 11, 114514 (2016)
  doi:10.1103/PhysRevD.93.114514
  [arXiv:1601.04027 [hep-lat]].

  
  \bibitem{Golterman:2016lsd} 
  M.~Golterman and Y.~Shamir,
  Phys.\ Rev.\ D {\bf 94}, no. 5, 054502 (2016)
  doi:10.1103/PhysRevD.94.054502
  [arXiv:1603.04575 [hep-ph]].
 
 \bibitem{Appelquist:2017wcg} 
  T.~Appelquist, J.~Ingoldby and M.~Piai,
  JHEP {\bf 1707}, 035 (2017)
  doi:10.1007/JHEP07(2017)035
  [arXiv:1702.04410 [hep-ph]].

  
    \bibitem{Cecile:2008kp} 
  D.~J.~Cecile and S.~Chandrasekharan,
  Phys.\ Rev.\ D {\bf 77}, 091501 (2008)
  doi:10.1103/PhysRevD.77.091501
  [arXiv:0801.3823 [hep-lat]].
  
  
\bibitem{Soto:2011ap} 
  J.~Soto, P.~Talavera, and J.~Tarrus,
  Nucl.\ Phys.\ B {\bf 866}, 270 (2013)
  [arXiv:1110.6156 [hep-ph]].

     
  \bibitem{Ametller:2014vba} 
  L.~Ametller and P.~Talavera,
  Phys.\ Rev.\ D {\bf 89}, no. 9, 096004 (2014)
  doi:10.1103/PhysRevD.89.096004
  [arXiv:1402.2649 [hep-ph]].
  
  \bibitem{Ametller:2015xva} 
  L.~Ametller and P.~Talavera,
  Phys.\ Rev.\ D {\bf 92}, 074008 (2015)
  doi:10.1103/PhysRevD.92.074008
  [arXiv:1504.06505 [hep-ph]].
  
  \bibitem{Hansen:2016fri} 
  M.~Hansen, K.~Lang¾ble and F.~Sannino,
  Phys.\ Rev.\ D {\bf 95}, no. 3, 036005 (2017)
  doi:10.1103/PhysRevD.95.036005
  [arXiv:1610.02904 [hep-ph]].
  
  \bibitem{GellMann:1960np}
    M.~Gell-Mann and M.~Levy,
  Nuovo Cim.\  {\bf 16}, 705 (1960).
  doi:10.1007/BF02859738
  
  \bibitem{Achasov:1994iu}
  N.~N.~Achasov and G.~N.~Shestakov,
  Phys.\ Rev.\ D {\bf 49} (1994) 5779.
  doi:10.1103/PhysRevD.49.5779
  
  \bibitem{Achasov:2007fz}
  N.~N.~Achasov and G.~N.~Shestakov,
  Phys.\ Rev.\ Lett.\  {\bf 99} (2007) 072001
  doi:10.1103/PhysRevLett.99.072001
  [arXiv:0704.2368 [hep-ph]].

  
  \bibitem{Bruns:2016} 
  P.~C.~Bruns,
  arXiv:1610.00119 [nucl-th].
  
  \bibitem{Pascalutsa:2003}
    V.~Pascalutsa and D.~R.~Phillips,
  Phys.\ Rev.\ C {\bf 67}, 055202 (2003)
  doi:10.1103/PhysRevC.67.055202
  [nucl-th/0212024].
  
  \bibitem{Pascalutsa:2006up} 
  V.~Pascalutsa, M.~Vanderhaeghen and S.~N.~Yang,
  Phys.\ Rept.\  {\bf 437}, 125 (2007)
  doi:10.1016/j.physrep.2006.09.006
  [hep-ph/0609004].
  
  \bibitem{McGovern:2012ew} 
  J.~A.~McGovern, D.~R.~Phillips and H.~W.~Griesshammer,
  Eur.\ Phys.\ J.\ A {\bf 49}, 12 (2013)
  doi:10.1140/epja/i2013-13012-1
  [arXiv:1210.4104 [nucl-th]].
  
  
\bibitem{Hoferichter:2011wk} 
  M.~Hoferichter, D.~R.~Phillips, and C.~Schat,
  Eur.\ Phys.\ J.\ C {\bf 71}, 1743 (2011)
  [arXiv:1106.4147 [hep-ph]].
    
  \bibitem{Ecker:1988te} 
  G.~Ecker, J.~Gasser, A.~Pich and E.~de Rafael,
  Nucl.\ Phys.\ B {\bf 321}, 311 (1989).
  doi:10.1016/0550-3213(89)90346-5
  
  \bibitem{Gasser:1982ap} 
  J.~Gasser and H.~Leutwyler,
  Phys.\ Rept.\  {\bf 87}, 77 (1982).
  doi:10.1016/0370-1573(82)90035-7
 
 \bibitem{Gasser:1983yg}
  J.~Gasser and H.~Leutwyler,
  Annals Phys.\  {\bf 158} (1984) 142.
  doi:10.1016/0003-4916(84)90242-2
 
  \bibitem{Scherer:2012xha} 
  S.~Scherer and M.~R.~Schindler,
  Lect.\ Notes Phys.\  {\bf 830}, pp.1 (2012).
  doi:10.1007/978-3-642-19254-8
  
\bibitem{Batley:2010} 
J.~R.~Batley {\it et al.}  [NA48-2 Collaboration],
   Eur.\ Phys.\ J.\  {\bf C70}, 635-657 (2010)
  
\bibitem{Taylor}J. R. Taylor, $Scattering$ $Theory$. John Wiley \& Sons, Inc, (1972).
  
\bibitem{Protopopescu:1972sp} 
  S.~D.~Protopopescu {\it et al.},
  Proc. Int. Conf. on Experimental Meson Spectroscopy, Philadelphia, Pa., Apr 28-29, 1972. N.Y., Amer. Inst. Phys., 1972. p. 17-58.
  
 
\bibitem{GarciaMartin:2011cn} 
  R.~Garcia-Martin, R.~Kaminski, J.~R.~Pelaez, J.~Ruiz de Elvira and F.~J.~Yndurain,
  Phys.\ Rev.\ D {\bf 83}, 074004 (2011)
  doi:10.1103/PhysRevD.83.074004
  [arXiv:1102.2183 [hep-ph]].
  
\bibitem{DonoghueGolowichHolstein}
J.~F.~Donoghue, E.~Golowich, B.~R.~Holstein, ``Dynamics of the Standard Model", p.~180 (Cambridge University Press, Cambridge, 1992). 
  
\bibitem{Donoghue:1988eea} 
  J.~F.~Donoghue, B.~R.~Holstein, and Y.~C.~Lin,
  Phys.\ Rev.\ D {\bf 37}, 2423 (1988).
 
\bibitem{Marsiske:1990hx} 
  H.~Marsiske {\it et al.}  [Crystal Ball Collaboration],
  Phys.\ Rev.\ D {\bf 41}, 3324 (1990).
   
  
\bibitem{Uehara:2008ep} 
  S.~Uehara {\it et al.} [Belle Collaboration],
  Phys.\ Rev.\ D {\bf 78}, 052004 (2008)
  doi:10.1103/PhysRevD.78.052004
  [arXiv:0805.3387 [hep-ex]].


\bibitem{Gasser:2005ud} 
  J.~Gasser, M.~A.~Ivanov, and M.~E.~Sainio,
  Nucl.\ Phys.\ B {\bf 728}, 31 (2005).
  [hep-ph/0506265].
  
  
\bibitem{GomezNicola:2007qj} 
  A.~Gomez Nicola, J.~R.~Pel\'aez and G.~Rios,
  Phys.\ Rev.\ D {\bf 77}, 056006 (2008)
  doi:10.1103/PhysRevD.77.056006
  [arXiv:0712.2763 [hep-ph]].
 
 \bibitem{Hanhart:2008mx} 
  C.~Hanhart, J.~R.~Pel\'aez and G.~Rios,
  Phys.\ Rev.\ Lett.\  {\bf 100}, 152001 (2008)
  doi:10.1103/PhysRevLett.100.152001
  [arXiv:0801.2871 [hep-ph]].

 
\bibitem{Doring:2016bdr} 
  M.~D\"oring, B.~Hu and M.~Mai,
  arXiv:1610.10070 [hep-lat].
 
 \bibitem{Ananthanarayan:2000ht} 
  B.~Ananthanarayan, G.~Colangelo, J.~Gasser and H.~Leutwyler,
  Phys.\ Rept.\  {\bf 353}, 207 (2001)
  doi:10.1016/S0370-1573(01)00009-6
  [hep-ph/0005297].
  
\bibitem{Achasov:2011}
 N.~N.~Achasov and A.~V.~Kiselev,
  Phys.\ Rev.\ D {\bf 83} (2011) 054008
  doi:10.1103/PhysRevD.83.054008
  [arXiv:1011.4446 [hep-ph]].
  
  \bibitem{Achasov:2012}
    N.~N.~Achasov and A.~V.~Kiselev,
  Phys.\ Rev.\ D {\bf 85} (2012) 094016
  doi:10.1103/PhysRevD.85.094016
  [arXiv:1201.6602 [hep-ph]].
  \end{thebibliography}

\end{document}